\documentclass[preprint]{aastex63}

\usepackage{natbib,amsmath, amssymb}
\usepackage{graphicx}
\usepackage{comment, url}

\shorttitle{Baryonic post-processing of $N$-body Simulations}
\shortauthors{Williams, Khan \& McQuinn}

\begin{document}
1
\title{Baryonic Post-Processing of $N$-body Simulations, with Application to Fast Radio Bursts}

\author{Ian Williams}
\affiliation{University of Washington, Department of Astronomy, 3910 15th Ave NE, Seattle, WA, 98195, USA}

\author{Adnan Khan}
\affiliation{University of Washington, Department of Physics, 3910 15th Ave NE, Seattle, WA, 98195, USA}

\author{Matthew McQuinn}
\affiliation{University of Washington, Department of Astronomy, 3910 15th Ave NE, Seattle, WA, 98195, USA}

\begin{abstract}

Where the cosmic baryons lie in and around galactic dark matter halos is only weakly constrained. We develop a method to quickly paint on models for their distribution.  Our approach uses the statistical advantages of $N$-body simulations, while painting on the profile of gas around individual halos in ways that can be motivated by semi-analytic models or zoom-in hydrodynamic simulations of galaxies. Possible applications of the algorithm include extragalactic dispersion measures to fast radio bursts (FRBs), the Sunyaev-Zeldovich effect, baryonic effects on weak lensing, and cosmic metal enrichment.  As an initial application, we use this tool to investigate how the baryonic profile of foreground galactic-mass halos affects the statistics of the dispersion measure (DM) towards cosmological FRBs.  
We show that the distribution of DM is sensitive to the distribution of baryons in galactic halos, with viable gas profile models having significantly different probability distributions for DM to a given redshift. We also investigate the requirements to statistically measure the circumgalactic electron profile for FRB analyses that stack DM with impact parameter to foreground galaxies, quantifying the size of the contaminating `two-halo' term from correlated systems and the number of FRBs for a high significance detection.  Publicly available Python modules implement our \textsf{CGMBrush} algorithm.

\end{abstract}


\section{Introduction}\label{sec:intro}
In a simpler universe in which the baryons could not cool, they would essentially trace the dark matter. However, our universe deviates far from this simple picture, as the baryons can cool and condense to high densities in dark matter halos. Yet, even adding cooling does not match the real complexity.  If there were only cooling, the baryons associated with each dark matter halo would have condensed into stars -- the Milky Way galaxy would have been ten times more massive \citep[e.g.][]{2010ApJ...717..379B, 2017ARA&A..55..389T}. Instead, galaxies are host for violent phenomena like supernovae and active galactic nuclei that prevent such massive galaxies from forming. These phenomena inject energy and ultimately redistribute the gas, possibly even to distances well outside their dark matter halos. In addition to learning the locations of the baryons at low redshifts, understanding this redistribution is likely key to such big questions as why some galaxies have stopped accreting gas and are becoming red and dead.  It is also important for realizing precision cosmology with weak lensing \citep{2018MNRAS.480.3962C, 2019JCAP...03..020S, 2020MNRAS.492.2285D, 2021MNRAS.502.5593O}. 

Most of the understanding of the distribution of baryons around halos is derived from observations of absorption lines against a bright background source, typically a quasar \citep{1969ApJ...156L..63B,2012ApJ...758L..41T, 2016ARA&A..54..313M}. 
This probe has revealed large column densities in atomic hydrogen and in many metal ions to impact parameters that extend 100-200\;kpc from their foreground galaxies.  However, while photoionization modeling has been used to constrain the amount of $10^4$K gas \citep[e.g.][]{2014ApJ...792....8W}, there is no direct way to use these observations to probe the bulk properties and radial profile of all the gas (most of which is likely at higher temperatures).

Two related probes of the circumgalactic medium (CGM) are the thermal and kinetic Sunyaev Zeldovich (SZ) effects \citep{1970Ap&SS...7....3S,1972CoASP...4..173S}, which measure respectively the inverse Compton heating and Doppler shifting of CMB photons when scattering off free electrons. When stacking thousands of halos, the thermal SZ has been used to measure the projected pressure profile of the gas in halos down to $\sim 10^{13}\,M_\odot$ \citep{2013A&A...557A..52P, 2015ApJ...808..151G, 2016PhRvD..93h2002S, 2021PhRvD.103f3514A}.  The kinetic SZ is a new frontier that has recently been used to measure the the electron profile in the $\sim 10^{13.5}M_\odot$ halos of massive luminous red galaxies \citep{2021PhRvD.103f3514A}. One limitation for both SZ effects is that the wide beam of CMB instruments makes it difficult to probe structure below the virial radius of galactic halos. Additionally, the kinetic SZ stacking requires redshifts to reconstruct the velocities, and the largest spectroscopic catalogues are of galaxies considerably more massive than the Milky Way. 

A new class of sources that provide an avenue for localizing the baryons is fast radio bursts (FRBs), which were rediscovered by \citet{2013Sci...341...53T}, confirming an earlier report \citep{2007Sci...318..777L}.   FRBs are a class of bright extra-galactic millisecond radio transients, which likely are sourced by magnetars \citep{2019A&ARv..27....4P,2019ARA&A..57..417C}.  There is a detectable FRB roughly once every minute somewhere on the sky, of which only a small fraction are currently detected. However, many wide-field instruments are now coming online and are expected to increase the number of detections by orders of magnitude \citep[see][for a review]{2021arXiv210710113P}. FRB observations directly provide the amount of electrons along each sightline as the signal traverses through the intervening medium. The dispersion measure (DM) of this signal is an observed quantity that measures the delay in the arrival time -- caused by the total electron column along the line of sight -- as a function of the frequency \citep{2014ApJ...780L..33M,2014ApJ...783L..35D,2020Natur.581..391M,2021arXiv210809881S, 2021arXiv210713692C, 2021arXiv210900386L}. The estimates of \citet{2014ApJ...780L..33M} suggest that FRBs may excel at constraining the gas profiles around galaxies.  


This paper develops a new technique to model the distribution of baryons, and is applicable to all these CGM observables.  As an initial application, we focus on how the baryonic profile of foreground halos affects the statistics of DM towards cosmological FRBs.  Our technique starts with an $N$-body simulation, which is the primary numerical method for modeling nonlinear structure formation in hundreds of megaparsec regions \citep{1983Natur.305..196C,1985ApJ...292..371D}. 
$N$-body simulations do not follow how gas cools and is redistributed by feedback. Hydrodynamic simulations of galaxy formation are the preferred method to capture this feedback physics \citep[e.g.][]{2017MNRAS.466.3810F,2020MNRAS.491.1190S,2020ApJ...898..148L,2019MNRAS.488.2549S,2020MNRAS.493.1461L}. Since the energy injection from supernovae and AGN is not resolved, these hydrodynamic simulations adopt varied prescriptions for injecting energy into the gas on resolved scales, around sites of star formation and AGN activity in the simulation \citep{2015ARA&A..53...51S,2016MNRAS.461L..32F}. 
The relatively small box of these hydrodynamic simulations means that they still struggle to adequately sample the cosmological density field \citep{2018ApJ...865..147Z}, an issue that is significantly mollified in the largest $N$-body simulations.  Thus, we opt for a hybrid approach that uses the statistical advantages of large $N$-body simulations, while painting on the distribution of gas around individual halos in ways that can be motivated by zoom-in hydrodynamic simulations of galaxies  (or semi-analytic models). 
Our method can further use CGM models based on observational measurements or even phenomenological parameterizations to understand what baryonic profiles the data can support.

It is useful to compare our baryon pasting method to other approaches. The `baryonification' method of \citet{2015JCAP...12..049S} adds a displacement owing to baryonic pressure and feedback in post-processing to the positions of particles in $N$-body simulations.  This method has been shown to reproduce the effect of baryons on the matter power spectrum seen in more complex hydrodynamic simulations and has been used to interpret data \citep{2022MNRAS.514.3802S}.  \citealt{2022_Osato} presents two post-processing methods to construct tSZ and kSZ maps from $N$-body outputs. The first method is halo-based, where a spherical profile is assigned to the positions of $\geq10^{13}M_\odot$ dark matter halos, ignoring the contribution of matter not associated with such halos. This approach has the thermal SZ effect in mind, where most of the contribution originates from massive halos. In their second method, they describe a more expensive particle-based method that includes both halo and field particles. The particle-based method handles halo particles by estimating the dark matter density from an $N$-body simulation and mapping this along with halo mass to the baryonic pressure and density obtained from an intra-cluster gas model.  Particles not associated with halos contribute to the density, but not the pressure.  A key difference between the latter approach (and also the \citealt{2015JCAP...12..049S} approach) and the \textsf{CGMBrush} algorithm presented here is that instead of processing every particle, we operate on a 2D projected density field, bin halos by mass, and use fast 2D convolutions to apply a gas model to the entire mass bin. This allows faster calculations of full 2D maps with less resources (an ordinary home computer suffices), and our implementation also adds flexibility in easily applying different models.\footnote{A fast emulator for the matter power spectrum for their flexible seven-parameter baryonic-profile model has been developed in \citealt{2021JCAP...12..046G} for the \citealt{2015JCAP...12..049S} method, allowing faster parameter space scans for this statistic.}

This paper is organized as follows. We discuss our \textsf{CGMBrush} method in \S~\ref{sec:methods}.  In \S~\ref{sec:application} we investigate the results of an application to Fast Radio Bursts. We summarize our conclusions in \S~\ref{sec:conclusions}. Appendix~\ref{sec:bolshoi} discusses the simulation utilized in our application,
Appendix~\ref{sec:justification} provides additional justification for our method and  Appendix~\ref{sec:resolution_analysis} discusses convergence testing.


\section{the \textsf{CGMBrush} method} \label{sec:methods}

We present a method that can be applied to any cosmological $N$-body simulation (also generalizable to hydrodynamic simulations). The primary requirement, at least for the FRB application presented here, is that the $N$-body simulation must resolve sub-Milky Way halo masses (down to $\sim 10^{11}M_\odot$) and be in a $\gtrsim 100$ comoving Mpc box that is large enough for a cosmologically representative sample of structures.  The method realizes where the baryons are located with a significant reduction in the computational cost relative to hydrodynamic simulations of galaxy formation at the expense of some self-consistency. 

This method adds to $N$-body simulations one thing they lack -- a prescription for the locations of the baryons, especially around galaxies where the baryons' dynamics are distinct from the purely gravitational dynamics of the dark matter owing to cooling and feedback. 
 Algorithmically, we extract the matter associated with dark matter halos from the density field. We then redistribute the 20\% baryon fraction of this matter using physically motivated models for the baryonic profiles around halos.  For the dark matter that lies outside of dark matter halos, this method assumes that there is a component of the gas that perfectly traces the dark matter with the cosmic ratio.  While this is almost certainly true when averaging over megaparsec scales and larger, it may not be on the outskirts of halos; as discussed in Appendix~\ref{sec:justification}, the clustering observables that are of most interest are fortunately not very sensitive to how one models the diffuse gas at halo outskirts.\footnote{We note that extrapolating a $c=10$ NFW profile indicates that 40\% of the halos mass lies between one and two virial radii, and detailed calculations show that this percentile is on-average higher for sub $M_*$ halos and lower for more massive halos \citep{10.1111/j.1365-2966.2008.14301.x}.  This means that even if this mass is also redistributed by feedback, it is a smaller fraction than the virialized mass.  The agreement of the observed Ly$\alpha$ forest column density distribution across redshift with simulations that use various feedback models suggests to us that around most halos gas outside the virial radius is not significantly redistributed \citep{2016ARA&A..54..313M}, although we note that around Luminous Red Galaxies kinetic SZ observations suggest that such redistribution may even happen out to three virial radii \citep{2021PhRvD.103f3514A}.}   

Large $N$-body simulations follow typically billions of particles, with the largest surpassing trillions. To reduce the size of outputs,  the particles are often projected onto an Eulerian grid with some interpolation kernel (such as ``nearest grid point'' or ``cloud-in-cell''; \citealt{2011ApJ...740..102K}). 
Our method has the advantage of working on the combination of these Eulerian grid outputs and a halo field we create from the simulations' catalogue of collapsed halos. Moreover, the method only requires the 2D projection of these grids, which allows our algorithm to achieve an enormous dynamic range. A fixed grid, while wasteful in many circumstances, is justified for the projected distribution of baryons as many halos intersect with essentially every sightline.


\subsection{Algorithm at a single redshift} \label{sec:algorithm_single}
\begin{figure}[htp]
    \centering
    \includegraphics[width=18cm]{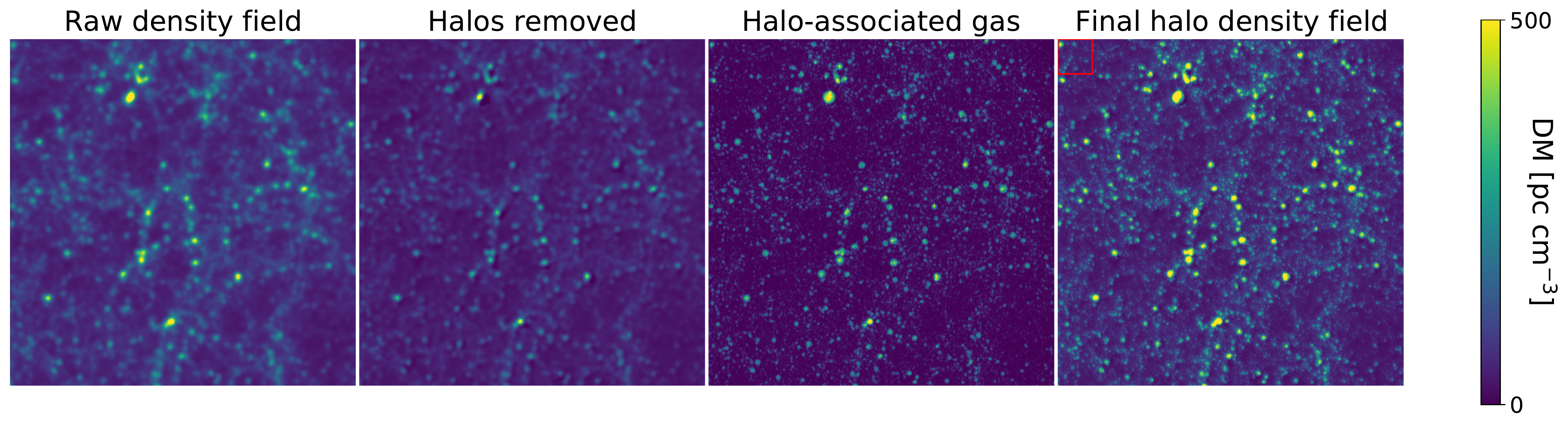}
    \includegraphics[width=18cm]{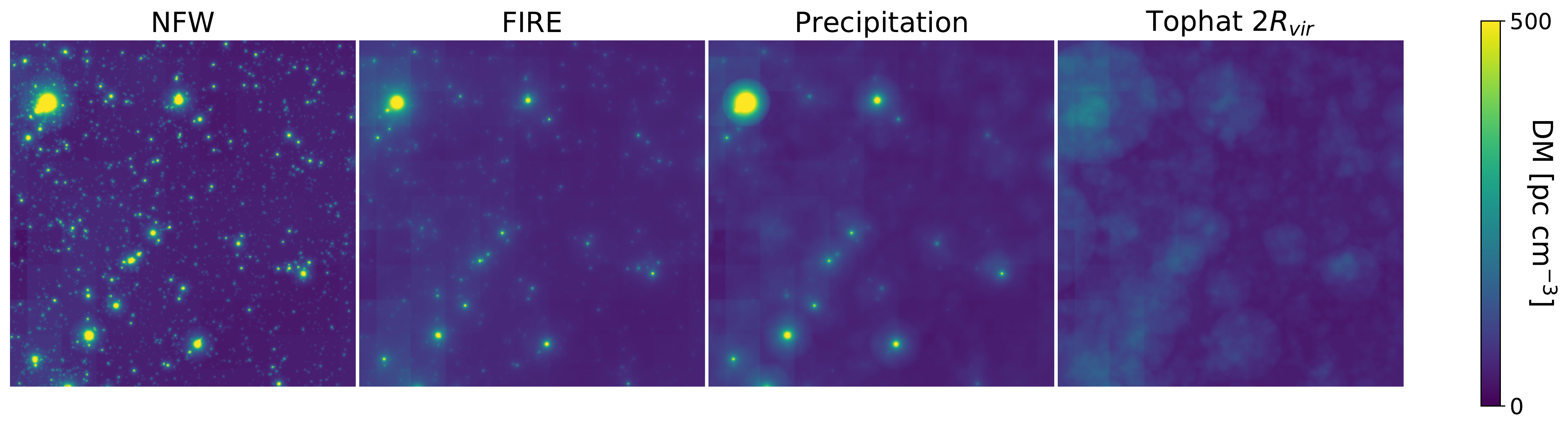}
    \caption{ Illustration of the \textsf{CGMBrush} method for different baryonic halo profiles applied to a $512^3$ grid of the $z=0$ snapshot of the Bolshoi $N$-body simulation. Each panel is projected over the simulation boxsize of $250h^{-1}$Mpc.  {\bf Top panels}: Shown is a $140\times140\,$Mpc$^2$ subregion of the full map of the dispersion measure (DM) -- the total electron column density along the line of sight -- through different stages of the algorithm.  
    The left-most panel shows the projection of the raw gridded density field from the simulation. From this density field, we subtract the template halo density field, and the result is shown in the second panel. The third panel shows the halo-associated gas density field using the $1 ~r_{\rm vir}$ top-hat profile that we wish to add back to the subtracted field, where a fine Eulerian grid of $N_f^2 =32,768^2$ is used (note that we find such a large $N_f$ is overkill). The right-most panel shows the projected final halo density field after the finely-gridded halo-associated gas density field is added back to the coarse subtracted field. {\bf Bottom panels}: A zoom-in on the $14\times 14\;$Mpc$^2$ region indicated by the red rectangle in the top-right panel, but shown for several other baryonic halo profiles. From left to right in the bottom panels, we show the final field when re-adding baryons with the NFW, FIRE, Precipitation, and $2~r_{\rm vir}$ tophat profiles.  These models are described in \S~\ref{sec:halo_profiles}, and we note that the NFW and tophat models are more of a toy character and not meant to be realistic.}
    \label{fig:procedure}
\end{figure}

We first describe how we generate a baryon field for a single redshift.   To start, the algorithm inputs a cosmological density field projected onto an Eulerian grid of $N^{3}$ numbers in a box of comoving size $L$.  The algorithm projects the 3D density grids to 2D grids with $N^2$ cells along a selected axis.  An example of this is shown using the $z=0$  snapshot of the Bolshoi simulation \citep{2011ApJ...740..102K} in the top left panel of Figure~\ref{fig:procedure} (refer to \S~\ref{sec:bolshoi} for details on this simulation). The acronym DM on the axes is because the gas column density is identical  (aside from redshift factors) to the dispersion measure -- an FRB observable considered in detail in later sections.

The next step in our algorithm is to remove the mass associated with dark matter halos from the Eulerian density field, which requires that we create a template halo density field. We create the template halo density field from the catalogue of collapsed halos in the simulation, where we require at a minimum their positions and halo masses. First, we divide the halos into $n_M$ equally spaced logarithmic mass bins between a minimum and maximum halo mass ($M_{\rm min}$ to $M_{\rm max}$) and assume all halos within a mass bin have the same halo profile. 
Then we calculate the virial radius of the halos for each mass bin, assuming the density profiles in a given mass bin specified by $M$ -- the average mass in the mass of bin -- is described by a single NFW profile \citep[e.g.][]{2002PhR...372....1C}
\begin{equation}
\rho(r|M) = \frac{\rho_{s}} {(r/r_{s}) [1 + (r/r_{s})^{2}]},
\label{NFW}
\end{equation}
parameterized by $r_{s}(M)= r_{\rm vir}/c$,  where $r_{\rm vir}$ and $c$ are given by
\begin{align}\label{eqn:virialrad}
    r_{\rm vir} = \left(\frac{3M} {4 \pi [18\pi^{2} - 82q(z) - 39q(z)^{2}] \rho_c }\right)^{1/3}; ~~~
    c(m,z) =  \frac{9}{1+z} \Big(\frac{M}{M_{*}}\Big)^{-0.13} ,
\end{align}
where $q(z) = {\Omega_{\Lambda}}/(\Omega_{m}(1+z)^{3} + \Omega_{\Lambda})$, $\rho_c = 3H^2/(8 \pi G)$, $M_{*} = 5 \times 10^{12} M_{\odot}$, and this form for the concentration parameter $c$ is taken from the numerical study of \citet{2000MNRAS.318.1144P}. The characteristic density, $\rho_{s}$, is set by the mass within the virial radius being $M$.

We then convolve the halo density profiles with the position of the centers of the halos (the physical location is provided by the simulation), yielding the halo density field template that we subtract off. Our algorithm does this projection using convolutions performed with fast Fourier transforms, which allows it to be very fast. One worry is that we have assumed a single halo profile, when actual relaxed halo profiles show significant scatter in their triaxiality and concentration \citep{2013ApJ...764L..31K}. This problem likely only has a minor effect on our results for reasons detailed in the next paragraph. 

The original projection of the $N$-body particles onto an Eulerian grid is effectively smoothing the density field, and this smoothing must be matched in this subtraction.  The simulator has choices in how this smoothing is performed.  In our application to the Bolshoi simulation, discussed later, the field is gridded by cloud-in-cell (CIC) interpolation and Gaussian smoothing with a standard deviation of one cell.  To create the template halo field, the algorithm puts down the halo profile at the position of each halo and also convolves this profile by the 2D projection of this smoothing.  A finer grid is used for this step than the $N^2$ projected simulation density outputs in order to avoid additional smoothing effects. In the application presented, we choose a $1024^2$ grid, slightly larger than our original density field. 
It then down-grids to the $N^2$ resolution of the simulation density outputs.  This final down-gridded halo density field is then used as a template for subtracting the halo contribution to the density field; the top left-middle panel of Figure~\ref{fig:procedure} shows the result of the halo field subtracted from the raw density field. Because the algorithm is designed so that the size of the coarse Eulerian grid cell ($1$~Mpc in our applications) is larger than the the virial radius of all but the largest halos, the method is relatively insensitive to the assumption of a single intrinsic halo profile in each mass bin (as given by eqn.~\ref{NFW}). The cost is that structures outside of halos (like sheets and filaments) that are smaller than the Eulerian grid cell are not resolved. Appendix \ref{sec:halomodel} argues that for our envisioned applications these structures are a subdominant source of fluctuations, and this is further quantified by our convergence test presented in Appendix~\ref{sec:resolution_analysis} for different sizes $N$ of the simulation's density grids.

 So far we have described how the method creates a halo field to subtract from the density field.  The most important step is to create a field that represents the halo-associated gas, which we aim to add back to the subtracted density field.  This step takes a profile for halo gas that can be motivated by simulations or models.  We describe some choices in \S~\ref{sec:halo_profiles}. To resolve the sub-halo structures when adding the halos' gas back with our desired profile, we increase the resolution of the 2D grid by a factor of $\eta$, so that our halo baryon grid and final total baryon density grid has resolution $N_f = \eta N$. The third panel from the right in the top row of Figure \ref{fig:procedure} shows the halo-associated gas grid for a final output grid of $N_f^2 =32,768^2$, assuming a simple 3D top-hat model for the halo-associated baryons that extends $1 \, r_{\rm vir}$ with all the baryons enclosed.  This large $N_f$ is overkill as all statistical quantities we consider are converged for considerably smaller grids, but it demonstrates how much resolution the 2D nature of our algorithm enables.  

Finally, we add the halo density field back to the raw density field from which the halos were previously removed to create a final grid of size $N_f^2$ (top right panel in Fig.~\ref{fig:procedure}).\footnote{In the implementation presented in this paper, we do not subtract the $\sim 5\%$ of baryons that form stars from the diffuse baryons, even though it does not contribute to the envision statistics such as regarding FRB DM. This omission is justified by the fact that removing this contribution will have a minor effect on any of our results, which are shaped by the large uncertainty in baryonic halo profiles.}\\


The summary of how the final baryonic halo field is generated for a single redshift is as follows:

\begin{enumerate}

    \item Start with a simulations density field on a 3D Eulerian grid, and project it over the ``line of sight" dimension to make the $N^3$ 3D grid an $N^2$ 2D grid (top left panel, Fig.~\ref{fig:procedure}).  The algorithm is designed to work where the $N^2$ grid does not resolve the virial radius of most halos, as will be the case for essentially any computable 3D Eulerian grid of a large $N$-body simulation.
    \item Create an $N^2$ template halo density field by convolving halo centers with a mean NFW halo profile in each halo mass bin and, then, by further applying the same smoothing/interpolation scheme as used to create the Eulerian density grid. (For our application to the Bolshoi simulation, CIC and Gaussian smoothing.)
    \item Subtract the template halo density field from the 2D density field to create a halo-subtracted baryon field  (top left-middle panel, Fig.~\ref{fig:procedure}).
    \item Add the baryons associated with halos back to the subtracted density field.  First, choose a baryonic profile at each halo mass (see \S~\ref{sec:halo_profiles} for potential models), and, next, convolve the halo centers with these profiles.  As halo gas can have structure on tens of kiloparsec scales, do this on a grid $N_f$ with much higher spatial resolution than the halo subtracted baryon field ($N_f \gg N$)   (top middle-right panel, Fig.~\ref{fig:procedure}).  
    \item Add the halo field back to the halo subtracted field to create the full model for the baryonic field on a $N_f^2$ grid (top right panel Fig.~\ref{fig:procedure}).

\end{enumerate}

\subsection{Halo Profiles}\label{sec:halo_profiles}

The previous section describes a procedure for modeling the distribution of gas around halos.  This allows one to study the effects of different baryonic profiles.
Here, we consider several models for the redistribution of gas in and around dark matter halos, some motivated by their simplicity, others by their analytical nature, and others still by sophisticated zoom-in simulations of the gas around halos.  Our code \textsf{CGMBrush} can also take a user specified profile, and supports combining multiple profiles together in a mass-dependent fashion.

In this paper we consider the following models:
\begin{description}
    \item[3D top hat] This is an unrealistic but instructive density model for the halos.  It assumes the closure density in baryons is distributed uniformly within a 3D sphere of radius $r$ around each halo center.  Perhaps the most motivated value is $r = r_{\rm vir}(M)$, but we also consider twice this extent with $r = 2r_{\rm vir}(M)$. 
     
     \item[NFW] The NFW density profile (eqn.~\ref{NFW}) describes the dark matter in halos rather than the gas \citep{1996ApJ...462..563N}.  However, it becomes a better approximation for the gas profile in the most massive galaxy clusters.  NFW models overestimate the concentration of the gas in smaller halos because cooling and feedback tend to redistribute the gas to larger radii \citep[e.g.][]{maller}. 
     
     \item[FIRE]  The FIRE profile is based on cosmological hydrodynamic ``zoom-in'' simulations with detailed sub-grid physics to account for feedback processes \citep{2018MNRAS.480..800H}. \citet{2019MNRAS.488.1248H} shows that the CGM density profiles in FIRE are well described by an $r^{-2}$ scaling.  Our method requires us to redistribute \emph{all} the baryons associated with these halos, whereas \citet{2019MNRAS.488.1248H} investigated the profile out to the virial radius, which misses baryons redistributed to further extents. Therefore, we assume an exponential cutoff at $r_{\rm max}$ that is determined so that the profile integrates to the total associated gas mass. Thus, our FIRE density profile is given by:
 \begin{equation}
 \rho = \rho_{0}\left(\frac{r}{r_{\rm vir}} \right)^{-2} \exp\left[-r/r_{\rm max} \right].
 \label{FIRE}
 \end{equation}
 \citet{2019MNRAS.488.1248H} measured the baryons contained within the virial radius of halos with $10^{10} M_\odot, 10^{11} M_\odot, 10^{12} M_\odot$ and $z=0.25$ using the FIRE cosmological simulations to be $0.1, 0.2$ and $0.3$. We use this to determine $\rho_0$ in eqn.~(\ref{FIRE}) and then set $r_{\rm max}$ to conserve the total mass.  We extrapolate the FIRE profile to higher halo masses using values for the baryon fraction motivated by $X$-ray observations ~\citep{2020MNRAS.491.4462D}.\footnote{The extrapolated fraction of baryons in CGM for $10^{13} M_\odot, 10^{14} M_\odot, 10^{15} M_\odot$ are $0.5, 0.8, 1$, respectively.}  
 \citet{2019MNRAS.488.1248H} finds a similar fraction of baryons within the virial radius for $z=0-2$, and so we assume no evolution in this fraction.  
 
 


     \item[Precipitation] This semi-analytic model specifies the density by assuming a threshold ratio of the cooling to dynamical time is satisfied \citep{sharma12, voit17}.  Idealized simulations show that exceeding this threshold results in dramatic cooling and fragmentation, that should source star formation and, hence, stellar feedback, and restore balance \citep{mccourt12}.  These models have found success at explaining various observations of gas in the Milky Way CGM \citep{2019ApJ...880..139V} as well as in other systems \citep{2019ApJ...880..139V, 2019ApJ...879L...1V}.  \citet{2019ApJ...880..139V} provides a fitting formula to their precipitation profile in physical units:
\begin{equation}
     n_{e}(r) = \left\{ \left [n_{1}\left (\frac{r}{1{\rm ~kpc}} \right)^{-\zeta_{1}} \right]^{-2} + \left[n_{2} \left(\frac{r}{100{\rm ~kpc}} \right)^{-\zeta_{2}} \right]^{-2} \right \} ^{-1/2}.
\label{preciptation-equation}
\end{equation}
The parameters of this model are given for a range of halo masses in \citet{2019ApJ...880..139V}.\footnote{Specifically, we use the \citet{2019ApJ...880..139V} models with $Z=0.3 Z_\odot$ and a cooling time to free fall time threshold of $10$.}
This density scales respectively at small and large radii as $ n_{e} \propto r^{-\zeta_{1}}$ with  $\zeta_{1} \approx 1.2$ and  $n_{e} \propto r^{-\zeta_{2}}$ with $\zeta_{2} \approx 2.3$.  The latter scaling is similar to the FIRE profile, although the fraction of baryons within the virial radius in the Precipitation model scales more strongly with halo mass than in FIRE. The normalizing factors $n_{1}$ and $n_{2}$ are interpolated from Table 1 in the appendix of \citet{2019ApJ...880..139V}, as are $\zeta_{1}$ and $\zeta_{2}$.  Just like with FIRE, we need to find a way to redistribution all the halo-associated gas. This density profile given by eqn.~(\ref{preciptation-equation}) is integrated out to $3r_{\rm vir}$.  If at some radius the total associated gas mass is exceeded, this becomes the maximum radius of the profile.  Otherwise, if by $3r_{\rm vir}$ some of the halo-associated gas mass is still not accounted for, this gas is then distributed in a tophat out to $3r_{\rm vir}$ (this tophat contains $98\%$ of the gas for a $10^{11}M_\odot$ halo, $78\%$ for a $10^{12}M_\odot$ halo, and no gas for a $10^{13}M_\odot$ halo at $z=0$).  The generalized result is that above $5\times10^{12} (1+z)^{-3/2} M_\odot$ there is no mass in the tophat. In order to prevent unrealistic densities in our extrapolation to large halos (as the tables in \citealt{2019ApJ...880..139V} only go to $10^{13}M_\odot$), we linearly extrapolate above this, and at $10^{14.1}M_\odot$ we transition to a NFW profile as the model starts to overshoot the NFW at this mass (there are few halos above this mass in our application to the Bolshoi simulation). This model is evolved in redshift from the $z=0$ fits in \citet{2019ApJ...880..139V} by keeping the profile fixed with constant virial temperature.\footnote{To extrapolate this model to higher redshifts from $z=0$ coefficients we use the fact that for an isothermal sphere mass profile, the dark matter density at fixed physical radius scales as $(1+z)M^{2/3}$, which means that fixing the dynamical time requires keeping  $(1+z)M^{2/3}$ constant.  However, if $(1+z)M^{2/3}$ is fixed, since $T_{\rm vir}$ has exactly the same scaling, this fixes the virial temperature and, hence, the cooling rate.  Thus, if we keep the same gas density profile as the $z=0$ density profiles of \citet{Voit_2018}, but identify this profile with a halo with mass such that $(1+z)M^{2/3}$ is fixed, this essentially fixes the ratio of cooling to dynamical times, which is what we desire to do to maintain the spirit of the model.  
}    


\end{description}

\subsection{Multiple redshifts and a light-travel image}

Most applications of our algorithm require capturing a light-travel image of the baryon field across a span in redshift. To create such an image with a periodic simulation box captured at discrete times, we follow the standard procedure used in many cosmological analyses of stacking shifted simulation snapshots, as well as interchanging which axis is the line-of sight \citep[e.g.][]{2014MNRAS.440.2610S}. The average redshift extent of our example box of size $250h^{-1}$~Mpc is $\Delta z \approx 0.1$ at $z \lesssim 1$. So to create a map out to higher redshifts than $z=0.1$, we must stack a number of boxes on top of each other.  To do so, we use the density and halo field snapshots closest to the mean redshift desired for the next box in the stack.\footnote{Since this stacking of a handful of boxes is a rather crude Riemann sum, to guarantee that the average over the box matches the expected DM(z) relation for our FRB application, we choose an effective redshift of each box in the stack to rescale its mean density to $\bar n_{e,0}(1+z_{\rm eff}^{(n)})^3$ and to assign to the $n^{th}$ snapshot, where $\bar n_{e,0}$ is the mean density today, so that the dispersion averages to the mean relation $\overline{DM}(z)$.  This requires
    $ z_{\rm eff}^{(n)} = ({\overline{\rm  DM}(z_{\rm max}^{(n)}) - \overline{\rm DM}(z_{\rm min}^{(n)})})/(L \bar n_{e,0}) - 1$,
where $z_{\rm min}^{(n)}$ and $z_{\rm max}^{(n)}$ are the redshifts for the edges of the $n^{\rm th}$ box of comoving size $L$ in the stack (calculated by converting the comoving size of the n and n+1 boxes to their corresponding redshifts).  This rescaling results in $z_{\rm eff}$ being just slightly different than the $z$ of the snapshot, but is preferable to the $10\%$ errors from $\overline{DM}(z)$ that result if we do not rescale for the applications presented in this paper.  Different weightings should be chosen for other observables.} A given snapshot can be selected multiple times if the cadence of simulation outputs is coarse.

\begin{figure}[htp]
    \centering
    \includegraphics[width=18cm]{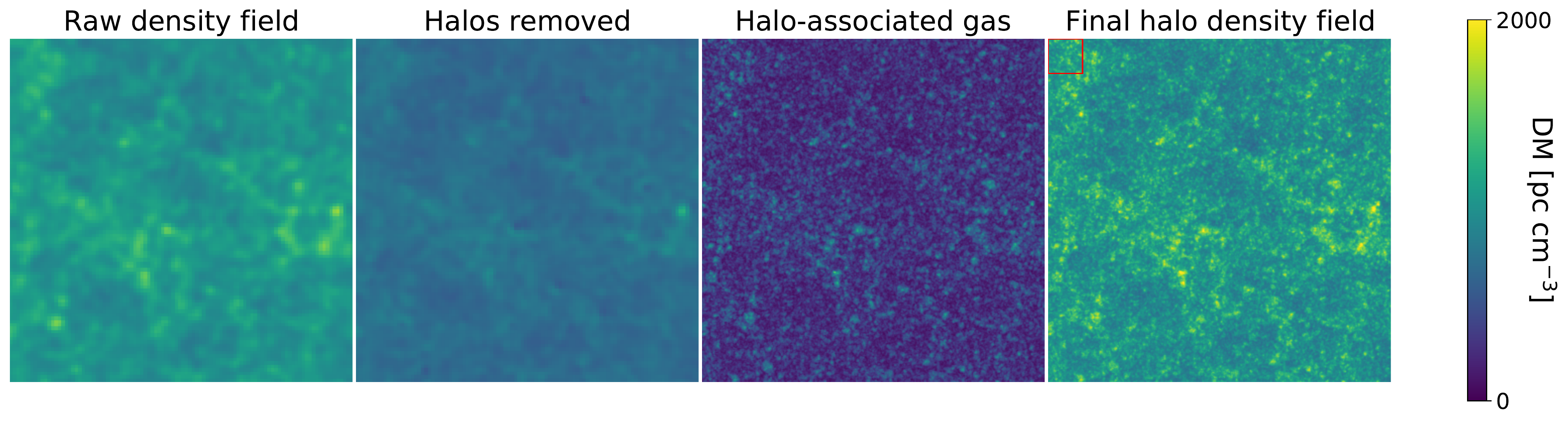}
    \includegraphics[width=18cm]{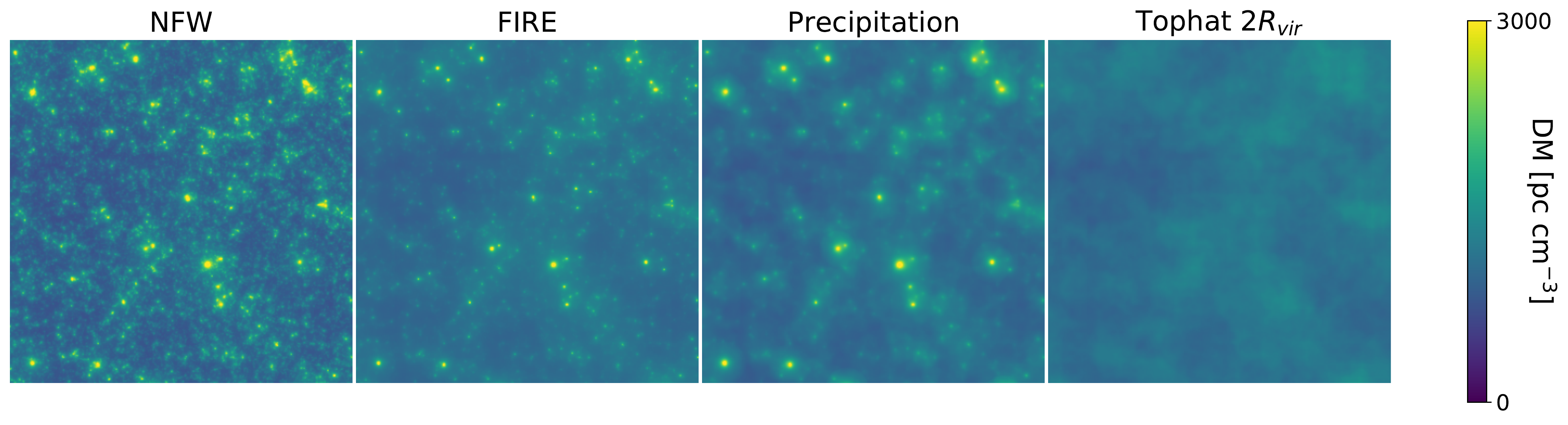}
    \caption{Illustration of the method for different baryonic profiles on a light-rectangle image to a redshift of $z=1$, where `light-rectangle' indicates that the projected coordinate follows the light travel delay. Nine boxes are projected over to construct the images. {\bf Top panels}: Shown is the DM in a $140\times140$Mpc$^2$ subregion through different stages of the algorithm. The left-most panel shows the projection of the raw gridded density field from the simulation. From this density field, we subtract the template halo density field, and the result is shown in the second panel. The third panel shows the halo-associated gas density field using the $1 ~r_{\rm vir}$ top-hat profile that we wish to add back to the subtracted field, where a fine Eulerian grid of $N_f^2 =8194^2$ is used. The right-most panel shows the projected final density field after the finely-gridded halo-associated gas density field is added back to the coarse subtracted field. {\bf Bottom panels}: A zoom-in on the $14\times14\;$Mpc$^2$ region indicated by the red rectangle in the top-right panel, but shown for other baryonic profiles. From left to right, we show the final light-rectangle field when re-adding baryons with NFW, FIRE, Precipitation, and $2~r_{\rm vir}$ tophat profiles.  These models are described in \S~\ref{sec:halo_profiles}, and we note that the NFW and tophat models are of a toy character and not meant to be realistic.}
    \label{fig:Procedure-z1}
\end{figure}

To create a light-travel image across multiple boxes, a randomization procedure is used to diminish artifacts from the same structures aligning in the stack.  For each box, the original density field is projected from 3D to 2D randomly along one of the three major axes.  Our normal procedure of removing and adding halos (\S~\ref{sec:algorithm_single}) is then applied to get the final density field for each box. The outputs are then randomly mirrored along each axis, and translated in the direction transverse to the line of sight by a vector of random direction and length sampled from $[0, \sqrt{2}L]$. The translation is performed in a manner that preserves the simulation's periodicity.  This stacking procedure assumes that correlations on scales larger than the box size can be neglected. The large box sizes our algorithm envisions (i.e. $250h^{-1}$ Mpc in the application we present), and the $\lesssim 1~$Mpc halo scale correlations of interest, suggest that this is a good approximation (see also Appendix~\ref{sec:justification}, which quantifies the contributions of fluctuations at $k\lesssim 2\pi/L$ to common variance statistics). The net result is that we have simulated the matter fields to high redshifts as shown in Figure~\ref{fig:Procedure-z1}.\footnote{Currently our maps are in position space instead of of angular space, which is convenient for the applications in this paper. To calculate angular statistics, like the angular power spectrum of the Sunyaev Zeldovich anisotropies or weak lensing, the algorithm needs to be generalized.  Angular maps are also interesting for our FRB application to understand DM correlations from more local structures such as the Local Group.} The rightmost image of the final halo density field is a $140$ Mpc region, and shows that our method is able to resolve structures and distribute baryons quite effectively without introducing any artifacts from redshift stacking. The zoomed in region of $14$ Mpc in the lower panel allows us to see the concentration of baryons for different profiles. As in Figure \ref{fig:procedure}, we see that the NFW profile is the most concentrated, followed by the FIRE, Precipitation, and then tophat profiles. We discuss the PDFs from these stacked fields in \S~\ref{sec:PDF_DM}.

\subsection{Computational Properties}

Our Python \textsf{CGMBrush} implementation of the algorithm allows an ordinary home computer to quickly re-paint baryons at high resolution. The (currently serial) implementation is capable of scaling $N_f$ to as a large a value as available memory supports. Our implementation requires $\sim4$ Gb RAM for $N_f=8,192$, scaling quadratically with $N_f$. The runtime for a single box at this resolution is about 7 minutes on a $\sim3$ GHz core, and scales roughly quadratically with $N_f$ and linearly with the number of boxes. 
Multiple realizations of the baryon field can be run as separate, parallel operations, provided enough memory is available. In a supercomputing environment, many profiles can be run at very high resolutions in a single day.

\section{Application of \textsf{CGMBrush} to FRBs}
\label{sec:application}


As an initial application of our algorithm, we explore how the baryon distribution around structures ranging from dwarf galaxies to rich clusters affect the dispersion measure (DM) of fast radio bursts (FRBs).

We apply our algorithm to data from the Bolshoi Simulation, described in Appendix~\ref{sec:bolshoi}. For all results presented below, \textsf{CGMBrush} calculations were made parameterized with $n_M = 60$ logarithmic mass bins between a minimum and maximum halo mass of $1\times10^{10} - 8.3\times10^{14 }~M_{\odot}$, resulting in $\Delta M \approx 0.21 M$. The upgridded resolution $N_f$ used are always specified with each result; we generally use $N_f=32,768$ for $z=0$ calculations and $N_f=8,192$ when computing quantities out to higher redshift (see Appendix~\ref{sec:resolution_analysis} for details on convergence).

\subsection{The probability distribution of DM$(z)$}\label{sec:PDF_DM}

\begin{figure}[htp]
    \centering
    \includegraphics[width=18cm]{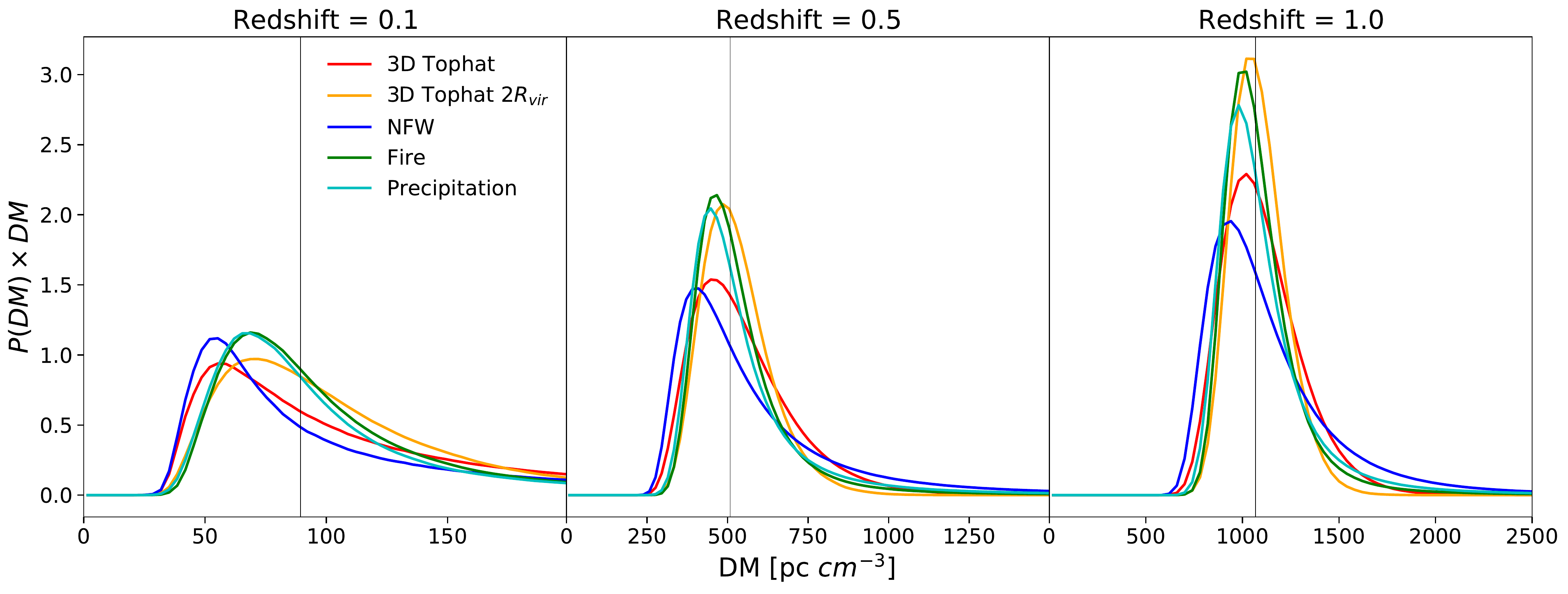}
    \caption{The PDF of DM for sightlines to the specified redshift for different baryonic profiles around halos. The different curves illustrate the dependence on different profiles including spherical tophat, FIRE, NFW, and Precipitation (the description of these profiles is given in \S~\ref{sec:halo_profiles}). The vertical line shows the mean DM to the given redshift.  These PDFs are calculated with $N_f=32,768$ resolution, although the Appendix~\ref{sec:resolution_analysis} demonstrates reasonable convergence on this statistic even for $N_f=1024$.}
    \label{fig:DM-redshift}
\end{figure}

\label{sec:PDM}

One of our main goals is to analyze the effect of different gas profiles on the probability distribution of DM to a given redshift. 
The DM is an observed quantity that measures the delay in the arrival time as a function of the frequency caused by the total electron column density along the line of sight. For FRBs at a redshift $z_s$:
\begin{equation}
    {\rm DM}(z_s, \boldsymbol{\hat{n}}) =  \int_{0}^{\chi(z_s)} \frac{\rho_e(z,\boldsymbol{\hat{n}})}{(1+z)^2} \,d\chi,
\end{equation}
where $d\chi = c \, dz/H(z)$ is the differential of the conformal distance, $\rho_e(z,\hat{n})$ is the electron number density at redshift $z$ in direction $\boldsymbol{\hat{n}}$, and $z_s$ is the source redshift.  We calculate statistics over the interval $z=0-1$.  Since the length of a single box is $250h^{-1}$Mpc, we stack nine different simulation snapshots to create the $z=1$ PDF. 

Figure \ref{fig:DM-redshift} shows the PDF of the cosmic distribution of DM to redshifts of 0.1, 0.5, and 1, respectively, for the profiles described above. The more diffuse the gas around halos or the rarer the halos that hold onto their gas, the more concentrated is the PDF. This trend arises because each sightline intersects a more statistically representative set of structures for the models where the baryonic profiles of the halos are more diffuse \citep{2014ApJ...780L..33M}.  

\begin{figure}[htp]
    \centering
    \includegraphics[width=18cm]{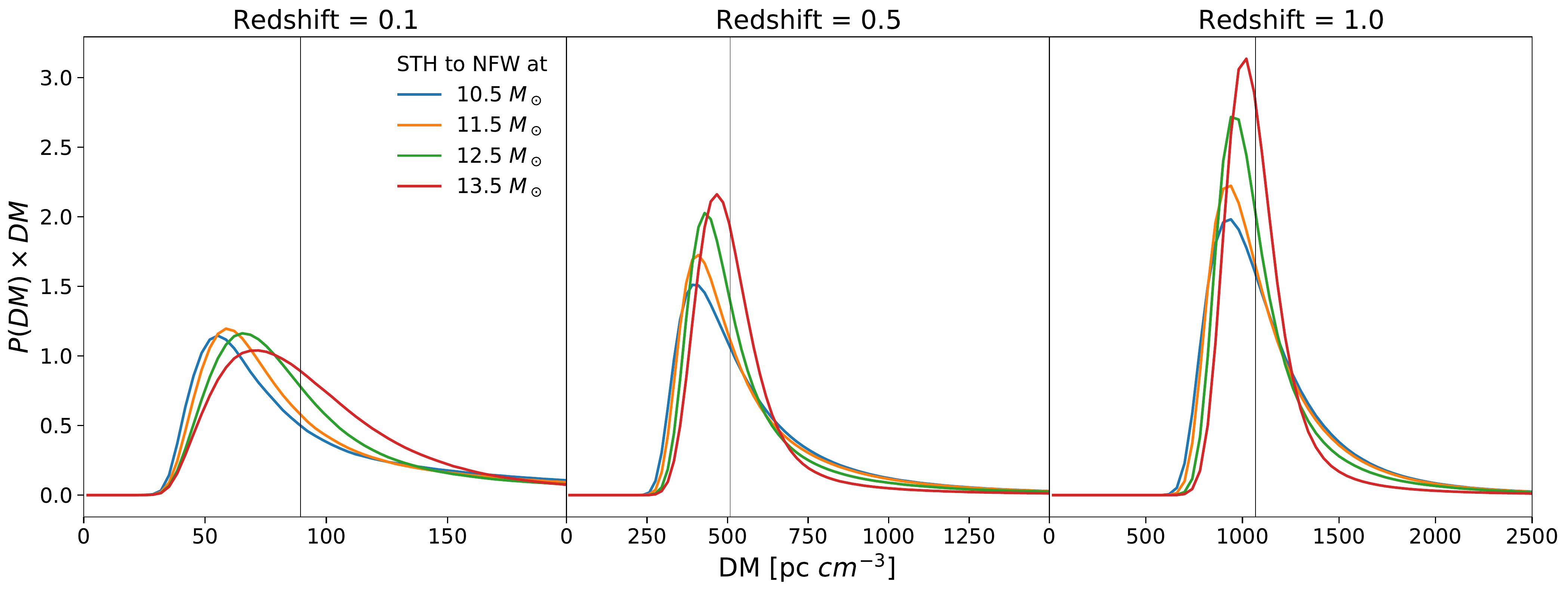}
    \caption{The PDF of DM for sightlines to the specified redshift using a toy model to understand how different halos contribute. In this model, we transition from a 2$r_{\rm vir}$ spherical tophat profile to an NFW profile above the mass thresholds listed in the legend.  The former profile we consider as the maximally evacuated case (where halos do not shape the width of the PDF) and the later NFW as the maximally concentrated. This transition has the effect of essentially eliminating the halo profile contribution to the PDF (the `1-halo' contribution).  This toy model illustrates where different mass halos likely contribute to the PDF and what mass scales need to be captured for convergence.  The vertical lines show the mean DM.  
    }

    \label{fig:STH_NFW_transition}
\end{figure}

Our algorithm enables us to investigate how different halo masses contribute to the PDF of DM.  Figure~\ref{fig:STH_NFW_transition} shows a series of PDFs in which we assume an extremely-puffy 2$\;r_{\rm vir}$ spherical tophat profile below the annotated mass threshold and dark-matter-tracing NFW profile above this threshold.  This transition has the effect of essentially eliminating the scatter contributed to the PDF from halos below the threshold mass, whereas the contribution of the more massive halos is being maximized since we are assuming the NFW profile, which acts as an upper bound on the gas concentration. The shift of the PDFs as a result of increasing the mass threshold shows that all mass scales matter from $10^{13.5}M_\odot$ down to $10^{10.5}M_\odot$ halos  (the leftward shift away from the mean vertical line indicates an increase in the scatter/variance of the PDF). The effect of the most massive halos on the PDF is significant, as observed by the shift in the curves from $10^{12.5}M_\odot$ to $10^{13.5}M_\odot$.  The most massive halos contribute to the high-DM tail. While none of the halo mass thresholds have a negligible effect, we observe a gradual convergence as the mass threshold decreases (so that the smaller halos are transitioning to the concentrated NFW profile).  Most of the width of the $10^{13.5}M_\odot$ case (in which most halos are evacuated) originates from the large-scale clustering of matter (rather than the Poisson fluctuations from sightlines intersecting halos) as there are few halos above this mass. Appendix~\ref{sec:halomodel} shows that matter clustering on $10-100\,$Mpc is never a negligible contribution to the PDF and can shape it in the more diffuse baryon profile models.

\subsection{DM versus impact parameter to galaxies} \label{sec:DM versus impact parameter to galaxies}

Ultimately, this baryon science from FRBs will not use the PDF of DM, but instead will stack foreground galaxies with different impact parameters to background FRBs to measure statistically the gas profile around halos.  Unlike the PDF of DM, stacking is unbiased by any DM intrinsic to the host system.  Such a stacking measurement was considered in \citet{2014ApJ...780L..33M}, but it has not been investigated using simulations.  A concern is the importance of correlating material in the stack, as stacking does not just measure the distribution of gas that is associated with the fraction of mass in baryons associated with each halo of mass $M$.  \textsf{CGMBrush} is useful understanding the size of this effect.

Unlike for $P({\rm DM})$, where we needed to trace through multiple redshifts, for the mean stacked profile of DM in a mass bin, we only need to consider a single box, as material at distances greater than the $250~\mathrm{Mpc}/h~$ simulation box -- to which our method is applied -- is weakly correlated with the halo used in the stack.  Material along each FRB sightline that is in the foreground or background will add to the variance of the stack, whose effect we can calculate from the $P({\rm DM})$ reported in the previous section.

\begin{figure}[htp] 
    \centering
    \includegraphics[width=12cm]{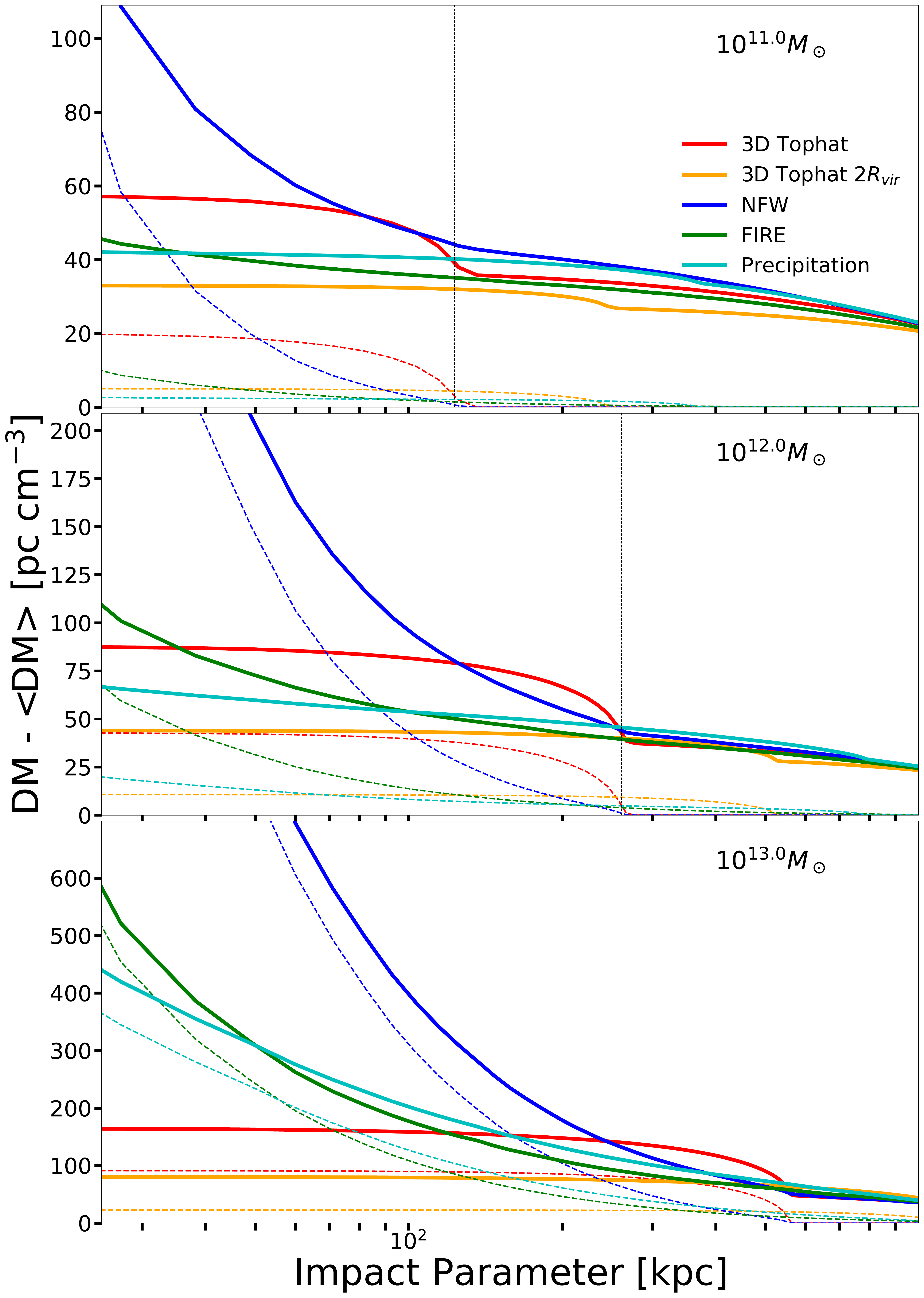}
    \caption{{\bf DM versus Impact Parameter}: The angular profile of DM for different halo gas profiles is shown in each panel, which can be measured by stacking FRB DM measurements by their impact parameter to foreground galaxies. The middle panel is for a Milky Way-like halo of $10^{12}M_{\odot}$, whereas the other panels consider galactic halos that are an order of magnitude more and less massive. The dashed lines show the DM for the baryonic profile we applied to the mass bin, and the solid lines show the DM extracted from the final density field produced by \textsf{CGMBrush}. The difference between the solid and dashed curves owes to the contribution of the gas correlated with halos that is not contained in the halo gas profile (i.e. the `two-halo' term).  The calculations are made at $N_f = 32,768$ resolution using the $N=512$ resolution initial Bolshoi density grid. } 
    \label{fig:DMvsRad_plot}
\end{figure}

We calculate the radial profile of halos in our final density field for different impact parameters as follows: for each halo within a mass bin, we trim out a grid around the halo from the final density field; then, we stack the trimmed grids of each halo in a mass bin on top of each other; from this stack of halos, we measure the density profile.  Figure~\ref{fig:DMvsRad_plot} shows the radial profiles of DM for $ 10^{11}~M_{\odot}$, $ 10^{12}~M_{\odot}$, and $ 10^{13}~M_{\odot}$ halos. The dashed lines show the DM of the baryonic profile applied to an isolated halo profile (the 1 halo term), and the solid lines show the DM of the final density field (both 1 and 2 halo terms). The vertical dashed line shows the virial radius of a halo of mass $M$. To the extent that there is no contribution of correlated material that lies outside of the halo, these two profiles should agree. The differences between the solid and dashed curves illustrates the contribution of all the matter correlated with the halo, the ``two-halo'' term. The two-halo term is much larger than the theoretical calculations by \citet{2014ApJ...780L..33M}; and generally the halo model has difficulty capturing the two halo term on such nonlinear scales.  In our calculations, the two-halo term is comparable to the one-halo term at all radii in the Milky Way-mass halos, and sets the trend outside the virial radius.   For smaller halos, the two-halo term is the dominant contribution at all radii, although the one-halo varies more quickly with impact parameter and so can still be distinguished.\footnote{It may be, especially towards lower mass halos, that the mapping from galaxy properties to halo mass cannot be approximated as one-to-one.  In this case, the interpretation of such a stacking experiment may be more complex.}


\begin{figure}[htp]
    \centering
    \includegraphics[width=12cm]{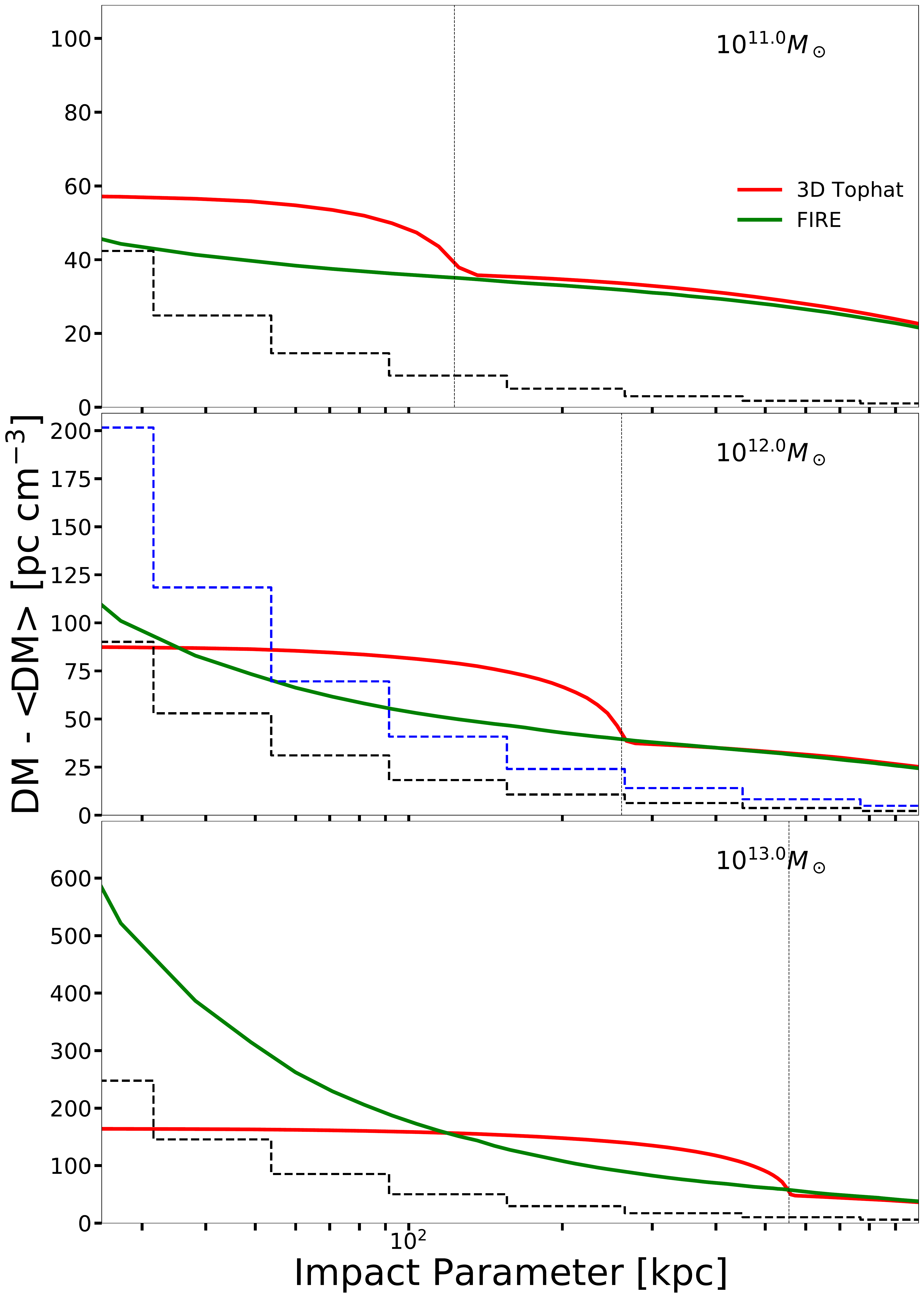}
    \caption{The angular profile of the dispersion measure for the 1\;$r_{\rm vir}$ 3D tophat and FIRE gas profile (solid curves) alongside our 1$\sigma$ error bar estimates for a mock stacking survey of localized FRBs (dashed piecewise lines).  This survey assumes 100 FRB localizations with a typical redshift of $z\approx 0.5$ plus models for the foreground incidence rate of different halo masses (see \S~\ref{sec:PDF_DM}). We use a variance of $\sigma_{\rm cosmic}=150$ for both profiles, as they were found to be very similar when computed from our $z=0.5$ PDFs (Fig.~\ref{sec:PDF_DM}). The blue dashed piecewise lines in the middle panel shows how the error increases if the FRB host system contributes scatter in DM of $\sigma_{\rm Host}= 300 $pc cm$^{-3}$, whereas the other piecewise lines are in the limit of the host galaxy being a subdominant contribution to the error. 
    }
    \label{fig:error_plot_DM}
\end{figure}

We next consider the sensitivity of a survey of $N_{\rm FRB}$ localized FRBs to measuring such a stack.  Such a survey's 1$\,\sigma$ error bar on DM$(b)$ in a radial bin is
\begin{align}
    \sigma(b) = \frac{\sqrt{\sigma_{\rm Cosmic}^{2} + \sigma_{\rm Host}^{2}} }{\sqrt{N_{\rm inter}(b)}}; ~~~ 
    N_{\rm inter} = {N_{\rm FRBs}} \times {N_{\rm halos}} \times P(b)
    \label{error_eqn}
\end{align}
where $N_{\rm inter}$ is the number of FRB sightlines that fall within $[b-\delta b/2,b+\delta b/2]$ for a  halo of mass $\sim M$.  In depends on $N_{\rm halos}$, the average number of halos intersected within a virial radius by each sightline to a redshift of $z \approx 0.5$ -- characteristic of a typical redshift of current FRB samples.  As read off from Figure 1 in \cite{2014ApJ...780L..33M}, $N_{\rm halos}$ equals approximately 4, 2, and 0.6 for halos with mass above $ 10^{11}~M_{\odot}$, $ 10^{12}~M_{\odot}$, and $ 10^{13}~M_{\odot}$, respectively. However, in our calculation we take more conservative estimates for halos intersected by each sightline to a redshift. First, \citet{2014ApJ...780L..33M} estimates are for intersecting halos above a mass threshold, whereas we are calculating the error for a mass bin; second, small halos are harder to detect. As such, for $10^{11}~M_{\odot}$ we reduce 4 to 1, and for $ 10^{12}~M_{\odot}$ we reduce 2 to 1. The $P(b)$ factor in $N_{\rm inter}$, gives the probability of the FRB signal passing through the gas for a given impact parameter bin at distance $b$ from the center of the halo and is given by $P(b) = (\pi (b+\delta b)^{2} - \pi b^{2})/(\pi r_{\rm vir}^{2}) $.

This error is plotted in Figure \ref{fig:error_plot_DM} alongside the $1~r_{\rm vir}$ tophat and FIRE profiles. The dashed black piece-wise line shows an estimate for the sensitivity to the mean DM for a tophat and FIRE profile for a sample of $N_{\rm FRB} = 100$ and $z\approx 0.5$ localized FRBs for halos of masses $ 10^{11}~M_{\odot}$, $ 10^{12}~M_{\odot}$, and $ 10^{13}~M_{\odot}$ in the top, middle, and bottom panel, respectively. These estimates use the quoted numbers for $z=0.5$ bursts as well as a $\sigma_{\rm Cosmic} = {150} $, the variance of the $P({\rm DM})$ in the FIRE model (the tophat was found to have very similar variance).  The significance that each profile can be detected in a radial bin is assessed by the ratio of the profile to the amplitude of the piece-wise line. The intrinsic contribution from the host galaxy is set to zero ($\sigma_{\rm Host} = 0$) in the black dashed piece-wise line. For this case and with 100 FRBs, we predict a stacking analysis will be sensitive to both models from $\sim 0.1~r_{\rm vir}$ outward, with this detection constraining the two-halo term outside $1~r_{\rm vir}$. 
Adding the $\sigma_{\rm Host} = 300$ pc cm$^{-3}$ as the intrinsic contribution from the host galaxy, which is likely an upper bound on the likely contribution to the error from the hosts \citep{2020Natur.581..391M}, the error increases significantly as shown by the dashed blue line, but the two profiles are still constrained down to a quarter or so of the virial radius.\footnote{Down-weighting the high-DM tail in a stack will reduce the effective variance, making our estimates conservative.  Since the cosmic contribution to the high-DM tail comes from sightlines through massive systems, such selection can also be done based on the proximity of sightlines to massive structures.  Selection can also use scattering, rotation measure, and the location within its host galaxy to excise hosts that likely have larger intrinsic DM.}

Thus, with as few as 100 FRBs and optical follow-up to find foreground galaxies, we predict that the DM profile can be constrained over an impact parameter of $\gtrsim 0.1 r_{\rm vir}$ for halo masses of $10^{10}-10^{13}~M_\odot$, with the sensitivity in detail depending on the true gas profile. While our method can measure the gas profile outside of the virial radius with such a sample, the two-halo term from correlating gas is generally dominant there, even for our gas profiles that distribute the most gas at $\gtrsim 1\, r_{\rm vir}$ radii.



\section{conclusions} \label{sec:conclusions}

Constraining the locations of the cosmic baryons is important for both understanding galaxy formation and achieving precision cosmology from large-scale structure surveys. We have developed a method to redistribute baryons around halos in $N$-body simulations and presented its application to FRBs.  The method is fast, only requiring convolutions in 2D, and it generates Eulerian 2D grids, which are easier to work with than adaptive outputs. A fixed grid, while wasteful for many applications, is more justified for the projected distribution of baryons as many halos intersect with essentially every sightline. The method enables modeling the baryons in much larger volumes and around smaller galaxies than can be achieved with modern cosmological hydrodynamic simulations, and it can be used to quickly survey potential models.  

The distribution of halo-associated baryons that are used can be informed by the results of semi-analytic models, zoom-in hydrodynamic simulations, or previous observations.  We implemented several such distributions in this study.  In addition to toy analytic models, we considered models for the baryonic profile motivated by the precipitation-limited models of \citet{2019ApJ...880..139V} and the FIRE galaxy formation simulations \citet{2019MNRAS.488.1248H}. Our publicly available Python package \textsf{CGMBrush} allows the user to create custom models.

By applying the \textsf{CGMBrush} method to the statistics of DM towards cosmological FRBs, we have shown that the DM is quite sensitive to the distribution of baryons in galactic halos, with plausible models producing significantly different probability distributions of DM to a given redshift. We have also investigated the projected profile that could be measured in a analysis that stacks FRBs based on their impact parameter to foreground galaxies of different types.  Our investigation included the first assessment of the effect of the contaminating matter from correlating systems, which we find complicates inferring the gas profile towards smaller mass systems. With as few as 100 FRBs that are localized to galaxies plus optical follow-up to find foreground galaxies, we predict that the DM profile can be well constrained over an impact parameter of $0.3-1 r_{\rm vir}$ for halo masses of $10^{11}-10^{13}~M_\odot,$ with the sensitivity in detail depending on the true gas profile. The gas profile outside of the virial radius would also be measured with such a sample, although the two-halo term from correlating gas is generally dominant for Milky Way-mass halos (and for even smaller halos, the two halo can be larger even at $<1\,  r_{\rm vir}$).    
Unlike many CGM observables that are most sensitive to group and cluster sized halos, the limiting factor for measuring the halo gas profile towards even dwarf galaxies is having a deep enough survey to identify these galaxies; if they can be identified, we conclude that their gas profile can be measured statistically with FRBs.  With the likely ``exponential'' increase of FRB detections that have localizations, with CHIME, HIRAX, CHORD and DSA forecasting thousands in the next few years \citep{2021arXiv210710113P, Amiri_2021, 10.1117/1.JATIS.8.1.011019, vanderlinde_keith_2019_3765414, Kocz_2019, 2019BAAS...51g.255H}, our estimates suggest these measurements have the potential to finally constrain the total gas profile around galaxies of different types.

Our algorithm is not just relevant to FRB DM science that this paper focuses on. With extensions that allow for magnetic fields and small-scale gas inhomogeneities, it could also be used to predict the rotation and scattering measures to FRBs.  Furthermore, our technique is relevant for modeling the thermal and kinetic Sunyaev-Zeldovich effects, which also probe the gas profiles around halos. Our  tool can straightforwardly be extended to generate fast, flexible models for the Sunyaev-Zeldovich angular power spectra.  Additionally, our method could be applied to understand the distribution of metals in the universe owing to galactic feedback. Indeed, a related post-processing of tophat enrichment profiles around simulated galaxies was used in \citet{2012MNRAS.420.1053B} to quantify the extent of enrichment -- a technique they devised because their suite of hydrodynamic simulations could not reproduce the scope of metal enrichment. Finally, our models can be used to model the  weak lensing power spectrum, where the uncertain gas distribution around halos is a major systematic for accurately estimating cosmological parameters. Current lensing analyses mainly use the predictions of hydrodynamic simulations to account for this uncertainty \citep{2018MNRAS.480.3962C, 2020MNRAS.498.2887F, 2021MNRAS.502.5593O}, whereas \textsf{CGMBrush} models have more versatility at the expense of some self-consistency.\\



\noindent We thank Zachary Hafen for helping with the prescription for the FIRE-inspired profile, G. Mark Voit for help in implementing the precipitation model,  and the CosmoSim Database for making the results of simulations publicly available. We acknowledge support from NSF award AST-2007012.

\bibliographystyle{apj}
\bibliography{references}

\appendix

\section{Bolshoi Simulation Specifications}
\label{sec:bolshoi}

Throughout this work, both as a demonstration of the algorithm and for our FRB application, we use the Bolshoi simulation \citep{2011ApJ...740..102K}, publicly available at {\it CosmoSim}.  
This $N$-body simulation meets our basic requirements: it resolves sub-Milky Way halos (to masses as small as $\sim 10^{10}M_\odot$), which we require for convergence on our statistics (Appendix~\ref{sec:resolution_analysis}),  and its boxsize $L = 250h^{-1}$Mpc is greater than our $100$ Mpc requirement, large enough for a representative sample of structures. There are $2048^3$ particles with mass $1.35 \times 10^8 ~M_{\odot}~h^{-1}$, and the simulation is run in $\Lambda$CDM cosmology \citep{2011ApJ...740..102K} with 
parameters $h=0.7$, $\Omega_m =0.27$, $\Omega_b = 0.0469$, $n=0.95$ and $\sigma_8 = 0.82$. The outputs are available for a large number of snapshots between a redshift of 0 and 17.

{\it CosmoSim} provides the Bolshoi simulation density field in two resolutions, $256^{3}$ and $512^{3}$.  These 3D grid are calculated from the simulation snapshots using cloud-in-cell interpolation and, then, a Gaussian convolution with a standard deviation of one cell. This smoothing must be matched in subtracting halos, as described in \S~\ref{sec:methods}. Since the higher resolution of $512^3$ is only provided at $z=0$, we use the $256^3$ grid for whenever $z>0$ is required, such as in \S~\ref{sec:PDF_DM}, and $512^2$ for $z=0$ calculations, such as the radial profiles in Figure~\ref{fig:procedure} and \S~\ref{sec:DM versus impact parameter to galaxies}.  We compare the $z=0$ results using the two different base grid resolutions in Appendix ~\ref{sec:resolution_analysis}, and show that there are relatively minor differences in our final results.

\section{The scales that contribute to the variance of the baryonic field}
\label{sec:justification}
\label{sec:halomodel}

In this appendix, we discuss what structures contribute to the variance of the cosmological baryon field.  We show that the variance of the dispersion measure mainly comes from gas in galactic dark matter halos and, to a lesser extent, gas that traces $\approx 100~$Mpc cosmic structures.  Structures that are somewhat less diffuse than halos are less important to model, which is fortunate because the method detailed in this paper is the least consistent in modeling them. 

\begin{figure}[htp]
    \centering
    \includegraphics[width=18cm]{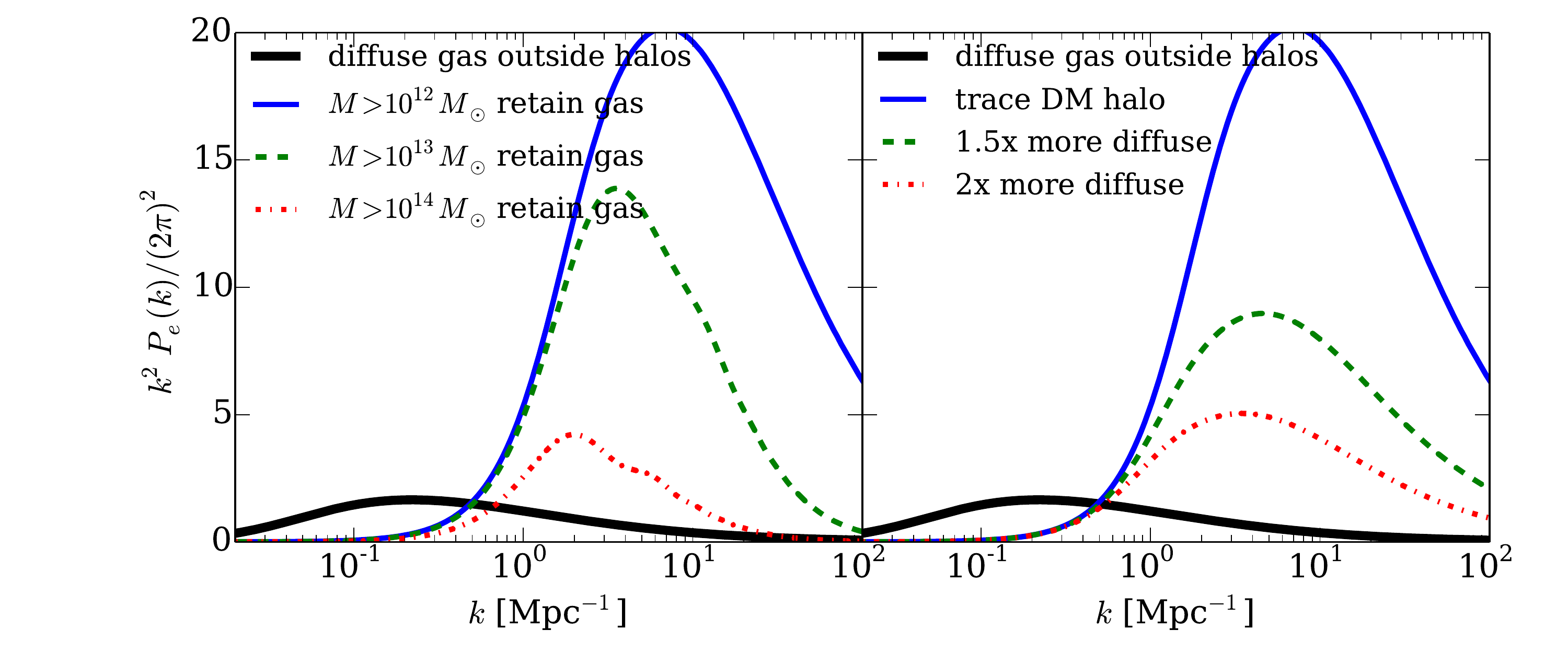}
    \caption{{\bf Left panel}: Halo model calculation for the $z=0.5$ contribution of the nonlinear power spectrum of baryons assuming a toy model where they trace the NFW halo profile above the specified halo mass and where halos are largely evacuated below this mass.  The colored curves show the one-halo term that depends on the profile, and the black curves show the two-halo term (which traces the large-scale matter field).  The area under these curves is proportional to the variance of DM (eqn.~\ref{eqn:sigmaDM}).  {\bf Right panel}: Considering only the model where $M>10^{12}~M_\odot$ halos retain their gas, but where the distribution of the gas is assumed to have an NFW profile that extends 1.5 or 2 times more in radius than the dark matter halo. This illustrates how the more diffuse the gas profile, the smaller the contribution of this one-halo term.}
    \label{fig:methodjustification}
\end{figure}

The left panel in Fig.~\ref{fig:methodjustification} shows $k^2 P_e/(2\pi)^2$ at $z=0.5$ for models where different halos retain their gas with the same NFW halo profile of the dark matter above the specified halo mass and that halos are largely evacuated below this mass.  These calculations use the halo model  \citep[e.g.][]{2002PhR...372....1C}.\footnote{Our calculation uses the same code as \citet{2014ApJ...780L..33M}, except here the Press-Schechter rather than Sheth-Tormen mass function is used. Press-Schechter somewhat overpredicts the abundance of rare massive halos and underpredicts less rare ones. } Here $k$ is the wavenumber and $P_e$ the electron overdensity power spectrum with the standard Fourier convention in cosmology such that the area under the curves is proportional to the variance of DM \citep{2014ApJ...780L..33M}:
\begin{equation}
 \sigma_{\rm DM}^2 \approx \int_0^z \frac{c dz}{H(z)} (1+z)^2 \bar n_{e,0}^2 \int d \log k  \left[ \frac{k^2 P_e(k, z)}{{(2\pi)^2}} \right],
 \label{eqn:sigmaDM}
\end{equation}
where $\bar n_{e,0}$ is the cosmic mean electron number density today.  The left panel shows that much of the variance potentially owes to the profiles in galactic mass halos.  A lesser contribution to the variance is from large-scale structure shown by the thick black curve. (This thick black curve is the halo model's `two-halo term'.)  Much of the area under this curve is at $2\pi/100$Mpc$^{-1}<k < 2\pi/10$Mpc$^{-1}$.

Large-scale structure becomes more important as the gas around halos becomes more diffuse. The right panel in Fig.~\ref{fig:methodjustification} considers the previous model where halos with mass satisfying $M>10^{12}~M_\odot$ retain their gas, but where an illustrative model for the distribution of the gas is assumed where it has an NFW profile that is dilated by a factor of $1$, $1.5$ and $2$ compared to the dark matter. This illustrates how the more diffuse the gas, the smaller contribution to the variance from sightlines intersecting individual halos, and the relatively larger contribution of the large-scale structure term.

In both panels, the large-scale structure term is calculated assuming the linear theory density power spectrum, as is traditional in the halo model.  The linear power spectrum is inaccurate of course on nonlinear wavenumbers of $k \sim 2\pi /(1-10)$Mpc$^{-1}$, but fortunately these scales are not a significant contribution to the variance.  That `two-halo correlations' are a somewhat small contribution to the power at $k \sim 2\pi /(1-10)$Mpc$^{-1}$ also holds for other baryonic clustering statistics and not just those most relevant for FRBs. Indeed, the unimportance of few Mpc scale structures is the reason why the halo model (which fails on these physical scales) is successful at predicting the nonlinear power spectrum at all wavenumbers. 

On $\sim 1~$Mpc scales the baryons that do not accrete onto dark matter halos may still be affected by feedback, which our \textsf{CGMBrush} algorithm is not modeling (since it only considers the gas mass that is associated with dark matter halos). While this may be a major deficiency of the algorithm, our justification is that regions somewhat outside of halos are less important for capturing many statistical measures of the cosmic baryon field.

\section{Resolution tests}
\label{sec:resolution_analysis}

\begin{figure}[htp]
    \centering
    \includegraphics[width=16cm]{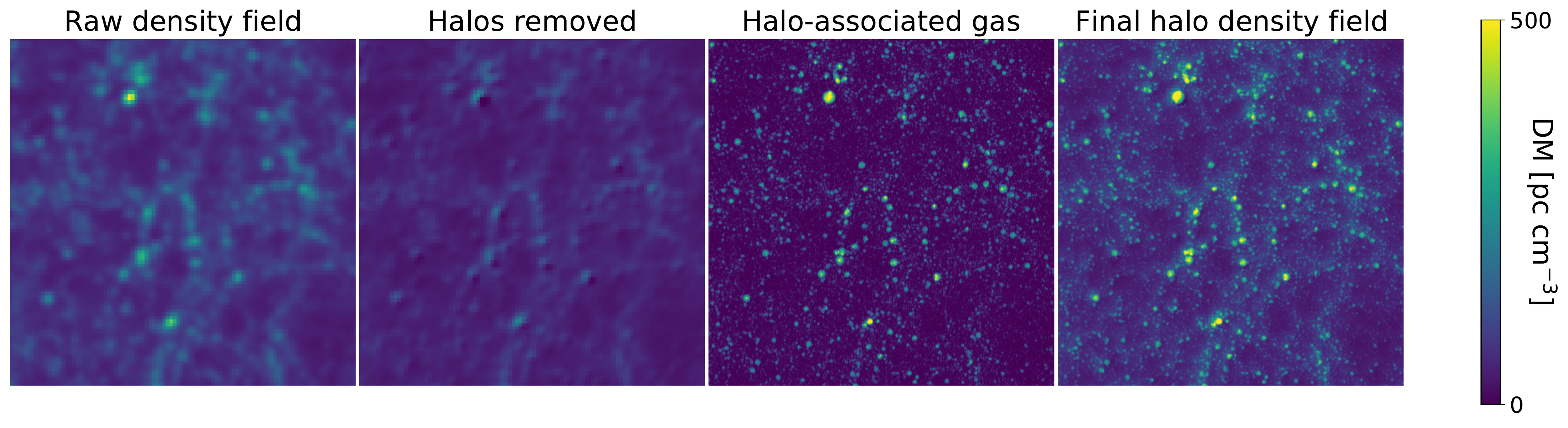}
    \includegraphics[width=16cm]{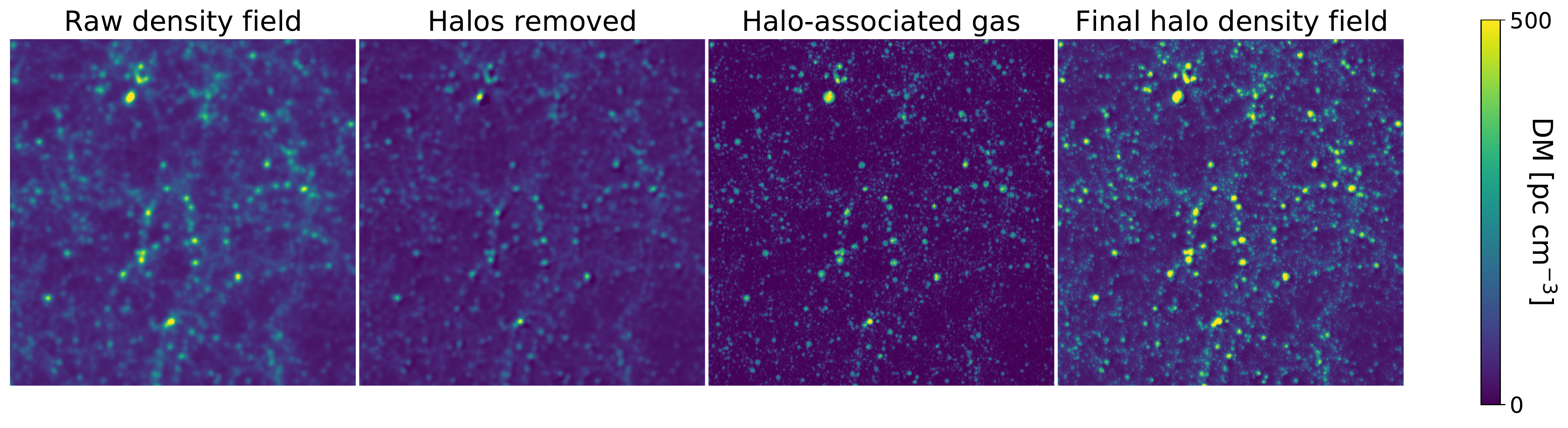}
    \includegraphics[width=14cm]{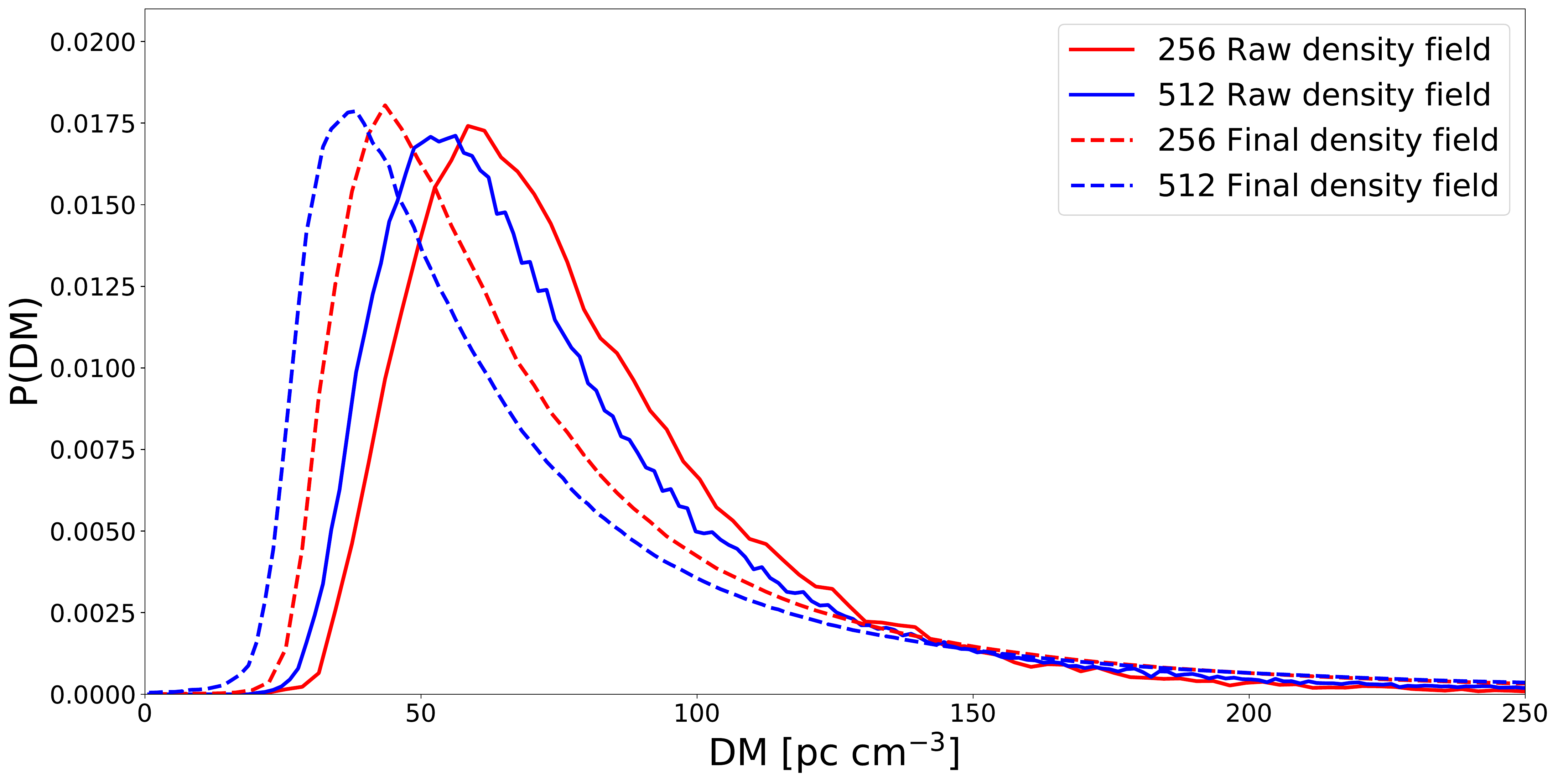}
    \caption{Illustration of method on two Eulerian base grid resolutions: the fiducial $N^2=256^2$ grid (top) and the higher resolution $N^2=512^2$ grid (middle). The images show the halo subtraction and addition through different stages of our method, applied to the $z=0$ snapshot of the Bolshoi $N$-body simulation. Each panel is projected over the simulation boxsize of $250h^{-1}$Mpc and zooms in on a subregion with transverse comoving size of $140$ Mpc. The left panels shows the raw gridded density field from the simulation. From this density field, we subtract the halo density field, and the result is shown in the second column. The third column shows the halo density field that we add back to the subtracted field, where a fine Eulerian grid of $N_f^2 =32,768^2$ is used. The last panel shows the net field after the finely-gridded halos are added back to the coarse subtracted field. In this case, we use our simple $1~r_{\rm vir}$ top-hat model, as the differences are smaller for more concentrated models. The bottom panel is a probability distribution of the gas column density, `DM', across the $250h^{-1}$ Mpc box. }
    \label{fig:256vs512resolution_comparison}
\end{figure}

\begin{figure}[htp] 
    \centering
    \includegraphics[width=11cm]{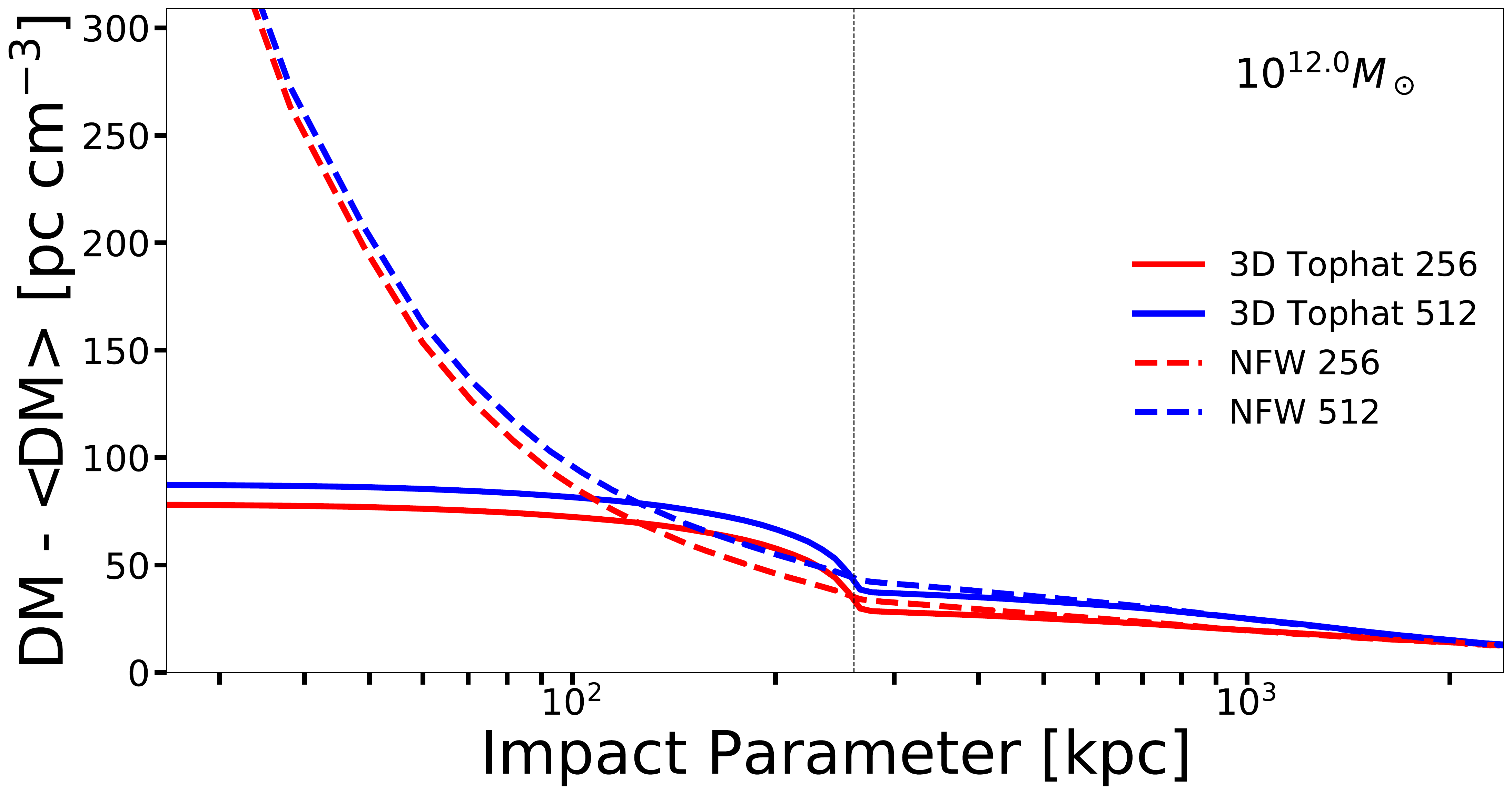}
    \caption{Comparison of the effect of using two different base grid resolutions to grid the simulation density field on the extracted radial profiles. The red lines use the fiducial $N^2=256^2$ grid of the Bolshoi simulation, and the blue lines use the higher resolution $N^2=512^2$ grid. The boost in DM from using the finer base grid is modest and owes to the two halo term being smoothed out on a scale of a cell.  The radial profile calculations use the $N^2=512^2$ grid in the main text. The upgridded resolution used for this comparison was $N_f^2=32,768^2$. }
    \label{fig:256vs512_dm_vs_r}
\end{figure}

We first consider how sensitive our results are to the density grid resolution of $N$-body simulation.
Figure~\ref{fig:256vs512resolution_comparison} shows the steps in our calculations for two different Eulerian grids for the $N$-body outputs. The top panel is for $N=256$ and the bottom panel for $N=512$. Visually, we can see that the differences are modest in the final algorithmic output in the rightmost panels. More quantitatively, we test convergence for the two base grid resolutions by creating a probability distribution function of the gas column density `DM' across before after our processing for the $1\; r_{\rm vir}$ tophat profile in the bottom of Figure~\ref{fig:256vs512resolution_comparison}. 
In Figure~\ref{fig:256vs512_dm_vs_r}, we examine the impact on the radial profiles discussed in Section~\ref{sec:DM versus impact parameter to galaxies}. While only the $10^{12}M_\odot$ bin is shown in Figure~\ref{fig:256vs512_dm_vs_r}, across $10^{11-13}~M_{\odot}$ halos, we find the differences are $\lesssim 20\%$, with convergence by 2~Mpc.  The differences owe to the two halo term being smoothed out on a scale of a cell, resulting in the $256$ falling below the the $512$.  We take these collective results to mean that our algorithm is reasonably converged, being only somewhat sensitive to the base gridding. The calculations in the main text use $N=256$ for high $z$ calculations and $N=512$ for $z=0$ calculations.

\begin{figure}[htp]
    \centering
    \includegraphics[width=16cm]{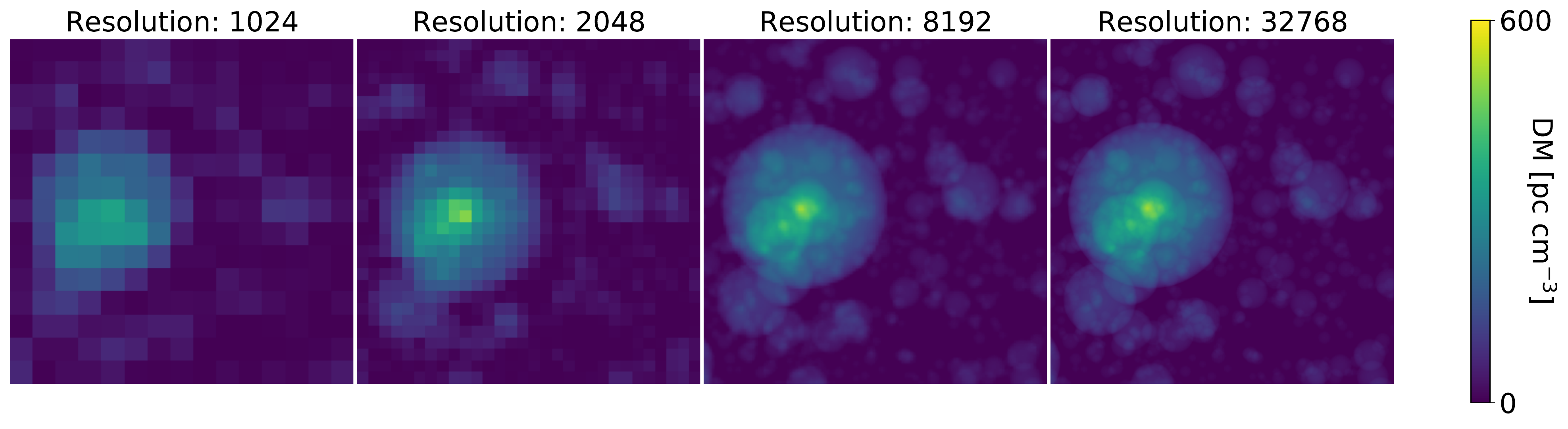}
    \vspace{.1cm}

    \includegraphics[width=13cm]{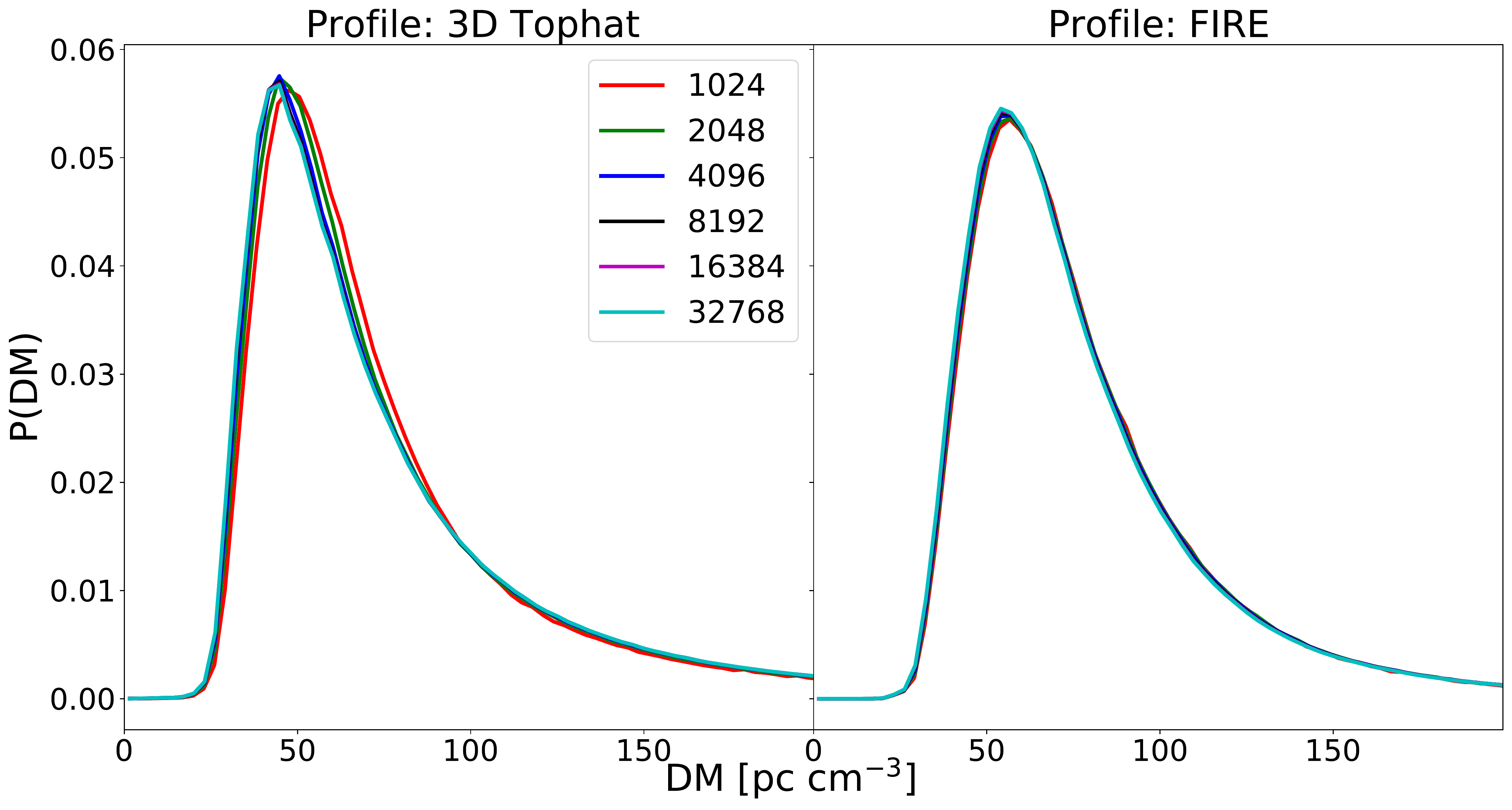}
    \caption{The images at the top show how the final projected density field converges when increasing the resolution of halo addition grid. Each image is computed by projecting the $z=0$ Bolshoi simulation box over $250h^{-1}$ Mpc and has a transverse comoving size of $5h^{-1}$~Mpc. The bottom panel shows how DM PDFs for the $1~r_{\rm vir}$ tophat (left) and FIRE (right) profiles tend towards convergence as we increase the resolution. Despite visual differences in the levels of structure resolved in the top panel, the PDF is not strongly dependent on the halo addition grid resolution. }
    \label{fig:implot_resolution_comparison}
\end{figure}

We also consider how sensitive our results are to the resolution of our up-sampled fine grid that we create to add halos back to the grid.  In the top panel of Figure~\ref{fig:implot_resolution_comparison}, where we have added the halos back by using a tophat profile (described in more detail in \S~\ref{sec:halo_profiles}), images of the final \textsf{CGMBrush} outputs are shown where as we increase the resolution to $\eta = 32$ in the third image, which corresponds to resolution of $N_f = 8192$, not only do we resolve the internal structure of larger halos of size $r_{\rm vir} \approx 1$ Mpc significantly, but also of the much smaller halos that were previously not well resolved as seen the left-most image in the panel with $\eta = 4$. Despite these visual differences in the images, the effect of increasing $\eta$ on the DM PDF is very small (see the bottom panels in Fig.~\ref{fig:implot_resolution_comparison}). All calculations in the main text use a resolution of 8,192 or higher.

\end{document}